\documentclass[twocolumn]{aastex702}

\usepackage{graphicx}
\usepackage{amsmath}
\usepackage{xcolor}
\usepackage{placeins}

\hypersetup{
    colorlinks=false, 
    pdfborder={0 0 0},  
}
\begin{document}

\title{On the Detectability and Measurement of Galactic Bars as a Function of Redshift}

\author{Yi-Xiao Lyu}
\affiliation{Kavli Institute for Astronomy and Astrophysics, Peking University, Beijing 100871, People's
Republic of China}
\affiliation{Department of Astronomy, School of Physics, Peking University, Beijing 100871, People's
Republic of China}
\email{yixiao_lv@stu.pku.edu.cn}
\author{Luis C. Ho}
\affiliation{Kavli Institute for Astronomy and Astrophysics, Peking University, Beijing 100871, People's
Republic of China}
\affiliation{Department of Astronomy, School of Physics, Peking University, Beijing 100871, People's
Republic of China}
\email{lho@pku.edu.cn}
\author{Zhao-Yu Li}
\affiliation{Department of Astronomy, School of Physics and Astronomy, Shanghai Jiao Tong University, 800 Dongchuan Road, Shanghai 200240, People's Republic of
China}
\affiliation{Key Laboratory for Particle Astrophysics and Cosmology (MOE)/Shanghai Key Laboratory for Particle Physics and Cosmology, Shanghai 200240, People's
Republic of China}
\email{lizy.astro@sjtu.edu.cn}
\author[0000-0001-5105-2837]{Ming-Yang Zhuang}
\affiliation{Department of Astronomy, University of Illinois Urbana-Champaign, Urbana, IL 61801, USA}
\email{mingyang@illinois.edu}

\begin{abstract}
The cosmic evolution of galactic bars offers key insights into the dynamical history of disk galaxies. However, studies of high-redshift bars are severely hampered by observational effects such as resolution degradation and surface brightness dimming. To quantify these biases and establish a robust methodology of bar detection and measurement for high-redshift galaxies, we develop an automated pipeline integrating isophotal ellipse fitting and Fourier decomposition. We benchmark its performance using a comprehensive set of mock observations generated by artificially redshifting high-quality images of local barred galaxies to match the background and resolution of HST ($z \approx 0.5-3$) and JWST ($z \approx 1-6$) observations. We find that applying standard bar detection criteria derived from local galaxies to high-redshift data results in a severe underestimation of the bar fraction, missing $>50\%$ of bars at $z>0.5$ in HST imaging and a similar fraction at $z>2$ in JWST imaging. Using adaptive, redshift-dependent criteria optimized via mock data achieves recovery rates above $60\%$ out to $z \approx 3$ for HST and $z \approx 6$ for JWST. The ellipse fitting method significantly outperforms Fourier decomposition in measurement reliability. While ellipse fitting systematically underestimates bar length by $\sim 4\%$--$36\%$ due to surface brightness dimming, this bias is consistent and correctable ($\sigma \approx 0.88$--$2.22$); in contrast, the Fourier method is prone to catastrophic failures ($\sigma \approx 2.40$--$6.14$) in the regime of low signal-to-noise ratio. We propose a standardized workflow that prioritizes mock-derived adaptive criteria and ellipse fitting, providing quantitative correction factors and empirical relations that enable future studies to recover intrinsic bar statistics from raw observations.
\end{abstract}


\section{Introduction} \label{sec:intro}

Bars are essential to elucidating the formation and evolution of disk galaxies \citep{kormendy_secular_2004}. Extensive optical and near-infrared (NIR) surveys show that most present-day disk galaxies host a bar. From early photographic work \citep{de_vaucouleurs_classification_1959} to modern imaging studies \citep{eskridge_frequency_2000,menendez-delmestre_near-infrared_2007}, the local fraction of disks with strong or weak bars converges to $\sim60\% \text{--} 70\%$, providing a robust $z\approx0$ benchmark for evolutionary studies. The independent high-resolution NIR analysis of \citet{knapen_subarcsecond_2000} found a bar fraction of $79\%\pm7.5\%$ among Seyfert galaxies and $59\%\pm9\%$ for a control sample of inactive galaxies, likewise supporting a high local incidence of bars. As shown by early observational and theoretical work \citep{simkin_nearby_1980, shlosman_bars_1989, heller_shlosman_1994} and by N-body and hydrodynamical simulations (e.g., \citealt{bournaud_gas_2002, bonoli_black_2016, spinoso_bar-driven_2017}) and numerous observational studies (e.g., \citealt{Ho_1997, sakamoto_bar-driven_1999, cheung_galaxy_2013}), bars can channel gas into the inner regions of galaxies, promote star formation to build bulge structures, potentially trigger central black hole activity, and drive quenching. Bars can repeatedly develop, disappear, and reassemble, affecting the evolution of disk galaxies over the course of their cosmic evolution (e.g., \citealt{hasan_chaotic_1990, bournaud_lifetime_2005}).

Understanding the development and evolution of bars depends heavily on the evolution of the bar fraction ($f_{\rm bar}$), the percentage of disk galaxies that are barred at different redshifts. However, different samples and procedures have produced varied results for this basic, but fundamental observational quantity. For instance, \citet{elmegreen_constant_2004} examined 186 disk galaxies observed using the Hubble Space Telescope (HST) and found a bar fraction of 23\% that was largely constant up to a redshift of 1, which they interpreted as evidence of sporadic bar disintegration. \citet{menendez-delmestre_near-infrared_2007}, using ellipticity and position angle criteria for bar identification based on isophotal analysis, reported a bar fraction of 59\% in a sample of 151 nearby spiral galaxies with ground-based NIR imaging. The bar fraction rose to 67\% when only the ellipticity attribute was taken into account, while the subset of disk galaxies with semi-major axes longer than 4 kpc constitutes only 31\% of the sources. A similarly high bar fraction was found in the NIR study of \citet{eskridge_frequency_2000}, while optical studies of $\sim$3500 Sloan Digital Sky Survey disk galaxies reported $f_{\rm bar}\simeq45\% \text{--} 55\%$ after controlling for inclination and resolution effects \citep{barazza_bars_2008,aguerri_population_2009}. Imaging using Spitzer at 3.6\,$\mu$m confirms this high incidence, measuring $f_{\rm bar}=66\%\pm4$\% for galaxies with stellar masses larger than $M_{\ast}\!\approx\!10^{9.5}\,M_{\odot}$ \citep{diaz-garcia_characterization_2016}. These local benchmarks (but see \citealt{elmegreen_constant_2004}) anchor all evolutionary comparisons that follow.

Beyond the local Universe, a series of HST surveys have traced the bar fraction out to \(z\approx3\). Early work on the Hubble Deep Field suggested a near absence of strong bars beyond \(z\gtrsim0.5\) \citep{abraham_morphologies_1994}, whereas the Groth Strip and Galaxy Evolution from Morphology and SEDs programs found $f_{\mathrm{bar}}\!\simeq\!25\% \text{--} 30\%$ at \(0.2<z<1.0\) after carefully matching brightness and size cuts to the $z\approx0$ benchmarks \citep{jogee_bar_2004}. Using the 2~deg$^{2}$ Cosmic Evolution Survey, \citet{sheth_evolution_2008} found a much steeper decline of bar fraction with increasing redshift, from $f_{\rm bar} = 65\%$ locally to \(\sim20\%\) at \(z = 0.84\). Morphological classifications from the Galaxy Zoo initiative \citep{melvin_galaxy_2014} and the Cosmic Assembly Near-infrared Deep Extragalactic Legacy Survey (CANDELS; \citealt{grogin_candels_2011}) broadly confirm a factor of $\sim$2 rise in $f_{\mathrm{bar}}$ from \(z\approx 1\) to 0, although \citet{kim_cosmic_2021} emphasize that the trend depends on stellar mass and bar strength selection. The significantly sharper resolution (factor of 2) and rest-frame optical and NIR wavelength coverage of the high-redshift galaxies observed by the James Webb Space Telescope (JWST) now resolve many compact bars that were missed by HST. In the Cosmic Evolution Early Release Science (CEERS; \citealt{finkelstein_ceers_2023}) and Public Release IMaging for Extragalactic Research (PRIMER; \citealt{Dunlop_2021}) pointings, the Near Infrared Camera (NIRCam) imaging detects bar fractions of $14\% - 18\%$ for massive ($M_{\ast}\!>\!10^{10}\,M_{\odot}$) disk galaxies at \(1<z<3\), about twice the value inferred from HST data in the same fields \citep{guo_first_2023,conte_jwst_2024}. The Euclid Quick Data Release (Q1) analysis now provides complementary wide-area constraints on the bar fraction in massive galaxies at $z<1$ \citep{euclid_collaboration_2026}. Taken together, the published bar fractions at \(z\approx1\) differ by roughly a factor of $2 \text{--} 3$ among different studies, underscoring how angular resolution, surface brightness dimming, and bandpass shifts systematically bias bar detection. Quantifying these effects is therefore essential before drawing any firm conclusions about bar formation and evolution over cosmic time---a key motivation for the analysis presented in this work.

Simulations provide an essential complement to observations by following bars in a fully cosmological context. The Evolution and Assembly of GaLaxies and their Environments (EAGLE) project predicts at \(z=0\) a population split among roughly 60\% unbarred, 20\% weakly barred, and 20\% strongly barred disks for $M_{\ast}\!=\!10^{10.6}-10^{11}\,M_{\odot}$ \citep{algorry_barred_2017}. In the Illustris-TNG suite, \citet{zhao_barred_2020} found that the bar fraction declines with redshift for \(0<z<1\) if galaxies are selected by a fixed stellar mass limit, yet remains \(\sim60\%\) when a redshift-dependent mass cut is imposed. Using the higher-resolution TNG50 run, \citet{rosas-guevara_evolution_2022} showed that the inferred trend also depends sensitively on the adopted threshold for physical bar size. High-resolution zoom-in suites such as the Auriga simulations reach $f_{\mathrm{bar}}\simeq50\%$ by $z\approx0.5$ and track bar growth back to $z\gtrsim3$ \citep{fragkoudi_chemodynamics_2020}. Overall, current cosmological simulations agree that massive disks can host bars early on, but the predicted redshift evolution of \(f_{\mathrm{bar}}\) varies by a factor of \(\sim2\) owing to differences in resolution, prescription for feedback processes, and sample selection. These discrepancies motivate an empirical calibration of observational biases, as undertaken in this work.

When studying the structural properties of bars, we often focus on basic observables such as the deprojected semi-major axis \(a_{\mathrm{bar}}\) and the ellipticity \(e_{\mathrm{bar}}=1-b/a\), with $b$ the semi-minor axis. Decades of work have shown that longer bars tend to reside in earlier type, more massive, and more luminous disk galaxies, although the correlations carry large intrinsic scatter \citep{kormendy_morphological_1979,elmegreen_properties_1985,ann_surface_1987,martin_quantitative_1995,erwin_how_2005,laurikainen_properties_2007,menendez-delmestre_near-infrared_2007,hoyle_galaxy_2011,diaz-garcia_characterization_2016,erwin_what_2019,lee_bar_2020}. Ellipticity, often combined with the Fourier $I_2/I_0$ amplitude or the normalized gravitational torque $Q_{b}$, is widely used as a proxy for bar ``strength'' and evolutionary stage \citep{laurikainen_comparison_2004,buta_fourier_2006,aguerri_population_2009}.

Bar identification and structural measurements rely on three main techniques: (1) ellipse fitting of isophotes, (2) Fourier decomposition of azimuthal light profiles, and (3) direct visual classification (e.g., \citealt{laurikainen_comparison_2004,menendez-delmestre_near-infrared_2007,melvin_galaxy_2014}). Two-dimensional (2D) decomposition, while more time-consuming, can yield more reliable bar sizes and strengths \citep{gao_optimal_2017}. However, even in the local Universe, different techniques can yield measurements that disagree by $\sim$20\% in both \(a_{\mathrm{bar}}\) and \(e_{\mathrm{bar}}\) for the same galaxy \citep{diaz-garcia_characterization_2016,gao_optimal_2017}. At higher redshift, point-spread function (PSF) blurring and surface brightness dimming further bias these quantities, systematically reducing both bar detectability and the accuracy of the measurements of $a_{\mathrm{bar}}$ and $e_{\mathrm{bar}}$ \citep{yu_redshifting_2023,liang_robustness_2024}. Published $f_{\mathrm{bar}}(z)$ trends therefore diverge, depending on sample selection and analysis pipeline \citep{sheth_evolution_2008,kim_cosmic_2021}.

To quantify and correct these observational biases, we create realistic mock images of well-resolved local barred galaxies. Following the redshifting recipe of \citet{yu_redshifting_2023}, originally pioneered by \citet{giavalisco_morphology_1996} and \citet{bergh_visibility_2002}, each local galaxy image is reprojected to a series of images at higher redshifts and observed with the corresponding HST and JWST filters. By applying ellipse fitting and Fourier analysis to both intrinsic and degraded images, we evaluate how PSF broadening, reduced signal-to-noise ratio, surface brightness dimming, and the minor mismatch in rest-frame bandpass that remains after choosing the nearest HST or JWST filter at each redshift collectively bias bar detection and the recovery of the basic structural parameters of bars. The empirical bias curves derived from these mocks will enable future surveys to correct high-redshift bar measurements in a self-consistent way.

Section~2 describes our parent sample and the redshifting pipeline that generates mock images. Section~3 describes the ellipse-fitting and Fourier analysis methods applied to the original and mock images. Section~4 presents bar recovery rates and measurement biases. Section~5 presents empirical corrections for ellipse-derived bar length and strength and discusses the astrophysical implications of these findings, caveats of the mock-based approach, limitations of 2D decomposition, and prospects for upcoming surveys. Finally, Section~6 summarizes our conclusions. We adopt AB magnitudes \citep{Oke_1983} and a flat \(\Lambda\)CDM cosmology with \(\Omega_{m}=0.27\), \(\Omega_{\Lambda}=0.73\), and \(h=0.73\).

\section{Mock Images of Barred and Unbarred Galaxies} \label{sec:style}

\subsection{Input Images from the Carnegie-Irvine Galaxy Survey}
\label{subsec:sample_selection}

Our mock analysis is based on the Carnegie--Irvine Galaxy Survey (CGS; \citealt{ho_carnegie-irvine_2011,li_carnegie-irvine_2011}), which provides high-quality optical ($BVRI$) images for a statistically complete sample of 605 bright ($B_T<12.9$ mag), southern ($\delta < 0\degr$) galaxies. We utilize the $R$-band images, which offer the best balance between depth (median surface brightness depth of 26.4~mag~arcsec$^{-2}$), spatial resolution (median seeing $\sim 1\farcs0$), and reduced sensitivity to dust extinction compared to bluer bands. A key resource for our study is the comprehensive structural analysis of the CGS sample performed by \citet{li_carnegie-irvine_2011}, who conducted detailed isophotal analysis to identify and measure galactic bars across the entire survey, resulting in a publicly available catalog of bar properties\footnote{CGS bar-properties database table.}. To calibrate our bar detection pipeline, we construct two distinct samples from this dataset: a sample of definitively barred galaxies and a control sample of galaxies classified as unbarred.

\begin{figure}[ht!]
\plotone{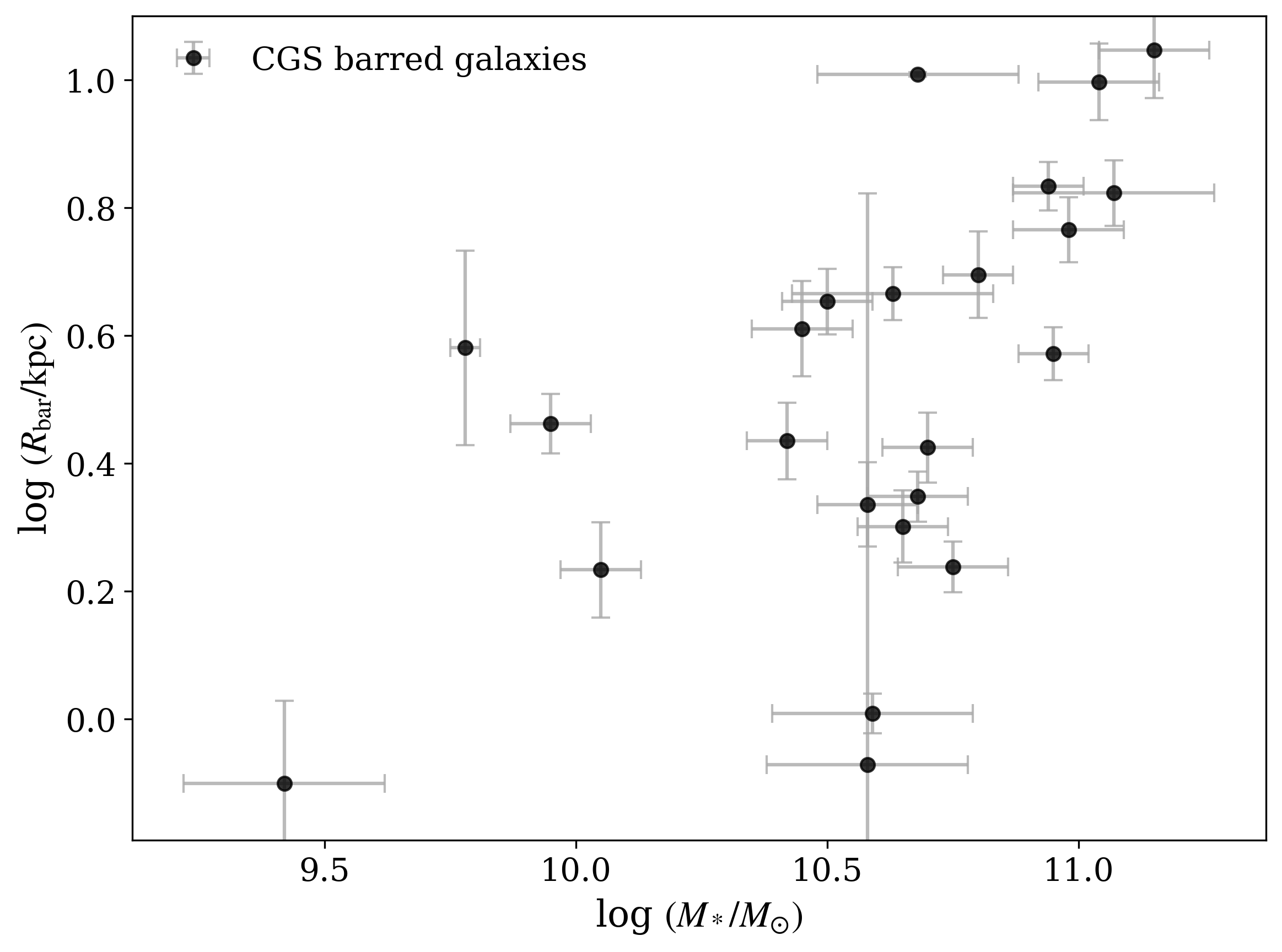}
\caption{Distribution of the selected CGS barred galaxies in stellar mass versus intrinsic bar length, using parameters from \citet{li_carnegie-irvine_2011}.}
\label{fig:sample}
\end{figure}

\begin{figure*}[ht!]
\centering
\includegraphics[width=0.70\textwidth]{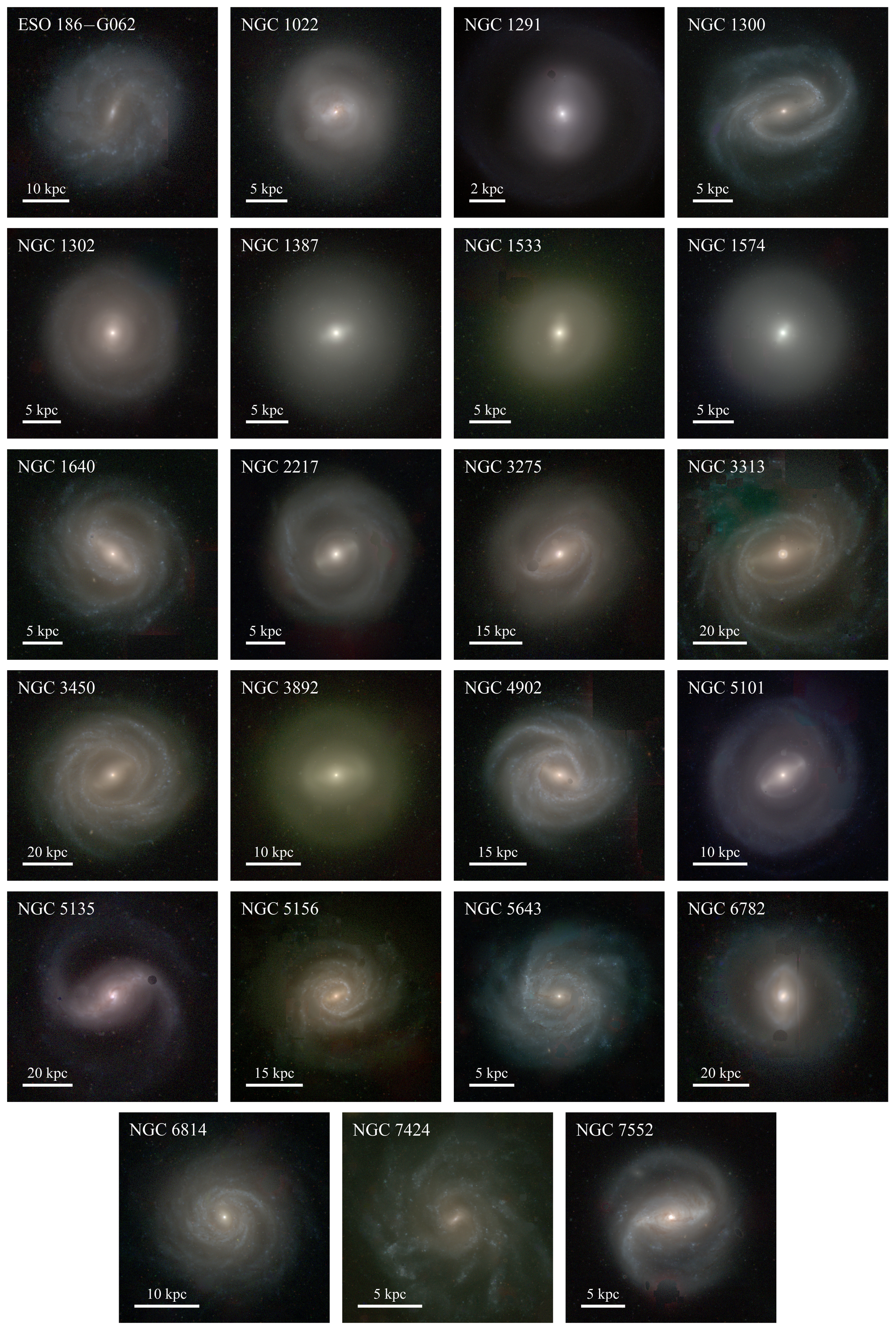} 
\caption{Color-composite, star-cleaned images of the selected barred galaxies from CGS \citep{ho_carnegie-irvine_2011}. North is up and east is to the left; a physical scale bar is shown in each panel.}
\label{fig:sample-image}
\end{figure*}

\paragraph{The barred galaxies}
We select galaxies identified as barred by \citet{li_carnegie-irvine_2011}, specifically those assigned the flag ``B'' (indicating ``barred''). To ensure that the sample is suitable for establishing an evolutionary baseline and minimizing projection effects, we restrict the sample to galaxies with relatively face-on orientation, by choosing disk inclination $i < 30\degr$ and stellar mass in the range $M_{\ast} = 10^{9} - 10^{11.5}\,M_{\odot}$ to cover the typical mass range of disk galaxies in high-redshift surveys such as CANDELS. This selection yields 23 barred galaxies, whose basic properties, including bar length ($R_{\rm bar}$, the semi-major axis of the bar) and bar strength [$e_{\rm bar}$, $(I_2/I_0)_{\rm bar}$] derived from \citet{li_carnegie-irvine_2011}, are summarized in Table~\ref{tab:cgs_sample_properties} and visualized in Figure~\ref{fig:sample}. Figure~\ref{fig:sample-image} displays their color-composite images.

\begin{figure*}[ht!]
\centering
\includegraphics[width=0.70\textwidth]{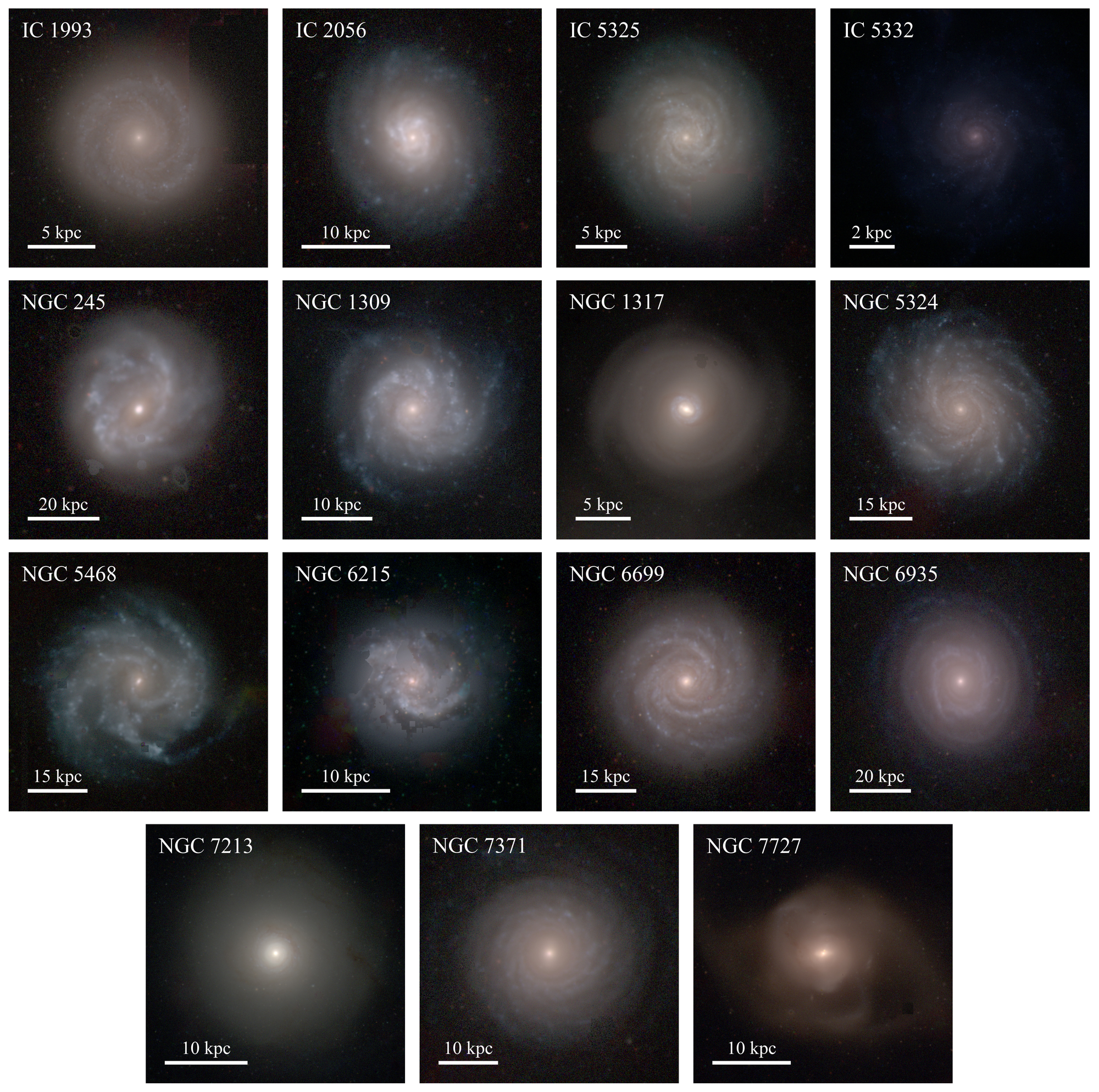} 
\caption{Color-composite, star-cleaned images of the control sample of 15 disk galaxies classified as unbarred in the CGS optical analysis. These galaxies were selected using the same mass and inclination criteria as the barred sample and are used to quantify the false-positive rate of our detection pipeline relative to the CGS optical classification. North is up and east is to the left. A physical scale bar is shown in each panel.}
\label{fig:unbarred_sample_image}
\end{figure*}

\paragraph{The unbarred control sample}
To evaluate the false-positive rate relative to the CGS optical classification, we select a control sample with flag ``N'' in the catalog of \citet{li_carnegie-irvine_2011} and apply the same inclination and mass cuts as for the barred sample. We analyze the CGS $R$-band images of these galaxies. We further visually inspect these candidates to exclude galaxies with potentially ambiguous optical morphological classifications and objects with severe foreground star contamination or dust obscuration that could interfere with isophotal analysis. This process results in a final control sample of 15 unbarred galaxies. Comparison with published Spitzer-based morphological classifications reveals that 10 galaxies in our control sample have entries in the Spitzer Survey of Stellar Structure in Galaxies (S$^4$G) or its extension, the Complete S$^4$G (CS$^4$G) \citep{buta_classical_2015,sanchez_alarcon_cs4g_2025}; an object-by-object comparison is presented in Appendix~\ref{app:control-audit} and Table~\ref{tab:control-audit}. We retain this optically defined control sample and evaluate the false-positive rate relative to the CGS classification. The longer-wavelength classifications show that this selection does not exclude weak, intermediate, or nuclear bars. Retaining weak NIR bars in this optically defined sample can inflate, instead of suppress, the apparent false-positive rate relative to the physical presence of a bar. Table~\ref{tab:unbarred_sample_properties} lists their physical properties, and Figure~\ref{fig:unbarred_sample_image} presents their color-composite images. These galaxies provide the negative-control set for our criteria optimization process (Section~\ref{subsubsec:optimization}).

\subsection{Method to Create Mock Images}
\label{subsec:mock_generation}
\begin{figure*}[ht!]
\centering
\includegraphics[width=\textwidth]{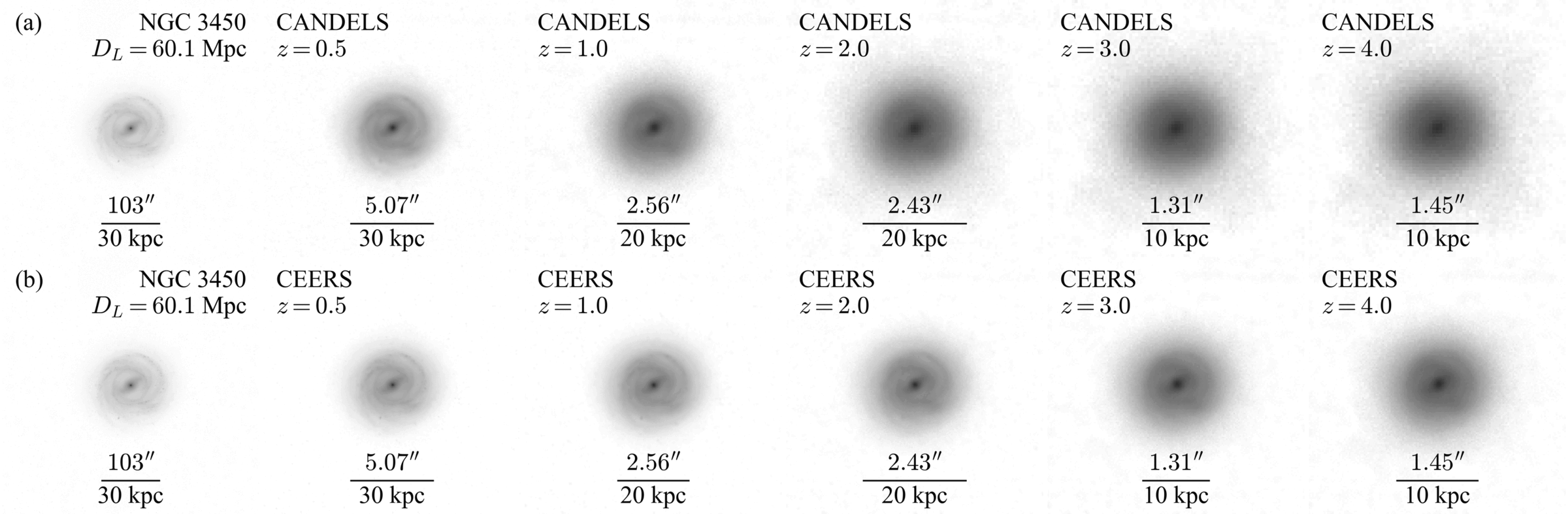}
\caption{Mock images of NGC\,3450 placed at higher redshifts, simulated for (a) HST/CANDELS in the EGS field, with mock images generated at $z = 0.5, 1.0, 2.0, 3.0,$ and $4.0$, and (b) JWST/CEERS, which provides deeper observations of the same EGS field, with mock images generated at $z = 0.5, 1.0, 2.0, 3.0,$ and $4.0$. The original CGS $R$-band image at $z \approx 0$ is shown on the left for comparison. The images are displayed with a logarithmic stretch to reduce saturation in the central regions. Each image is scaled independently for display and therefore does not provide a direct comparison of surface brightness between redshifts. These two sets of simulations represent observations with different resolutions, depths, and instrumental properties.}
\label{fig:3}
\end{figure*}

We generated mock images based on the selected galaxies to investigate the impact of redshift effects on the measurement of galaxy bar structures. The mock images were created by adjusting various parameters, including brightness, signal-to-noise ratio, resolution, and others, to simulate the appearance of the galaxies at higher redshifts. In addition to distance factors, the luminosities and sizes of the galaxies were also modeled to reflect their evolution over cosmic time. This evolution modeling was informed by established trends observed in deep survey fields, such as the Extended Groth Strip (EGS; \citealt{grogin_candels_2011, koekemoer_candels_2011}), which is the target field for both our CANDELS and CEERS simulations. The adjustments were made according to the galaxy's stellar mass. Figure~\ref{fig:3} shows the original image of NGC\,3450 alongside its mock images at different redshifts, simulated as if observed by the HST CANDELS and the JWST CEERS surveys. The generation of these mock images involved the following steps.

To ensure that the physical size of galaxies matches at different redshifts, we calculate the binning factor, taking into account the angular size and luminosity distance. When shifting an object from an initial redshift $z_i$ to a final redshift $z_f$ ($z_i \leq z_f$), its physical size remains constant. The angular size $a$ is related to the luminosity distance $D_L$ through

\begin{equation}
a_i \frac{D_{L,i}}{(1 + z_i)^2} = a_f \frac{D_{L,f}}{(1 + z_f)^2}.
\end{equation}

\noindent
For two imaging surveys with pixel scales $p_i$ and $p_f$, define $B=n_i/n_f$, where $n_i=a_i/p_i$ and $n_f=a_f/p_f$ are the numbers of pixels across the same physical size. The geometric resampling factor is

\begin{equation}
B = \frac{p_f}{p_i}\frac{D_{L,f}}{D_{L,i}}\left(\frac{1+z_i}{1+z_f}\right)^2.
\end{equation}

The output pixel span is $n_f=n_i/B$; thus $B>1$ reduces the number of pixels across the source. Utilizing the calculated binning factor, we bin the pixels in the object image and its corresponding PSF image to account for the instrumentation effects of real observations at the desired redshift. Before this step, we ensure that the full width at half maximum (FWHM) of the initial PSF image in physical units is smaller than that of the final PSF image to avoid introducing additional signals. We employ a flux-conserving algorithm provided by the Python package \texttt{reproject} \citep{robitaille_reproject_2020} to perform pixel resampling. This method treats sky pixels as four-sided spherical polygons and accurately calculates the intersection between them. For both the HST/CANDELS and JWST/CEERS simulations, we follow the PSF-matching procedure adopted by \citet{sun_structure_2024}. We compute the PSF-matching kernel using the \texttt{Photutils} routine \texttt{create\_matching\_kernel} \citep{bradley_astropyphotutils_2022}. Finally, we convolve the initial image after binning with the PSF-matching kernel to obtain the PSF-matched galaxy image.

\begin{figure}[ht!]
\centering
\includegraphics[width=8.5cm]{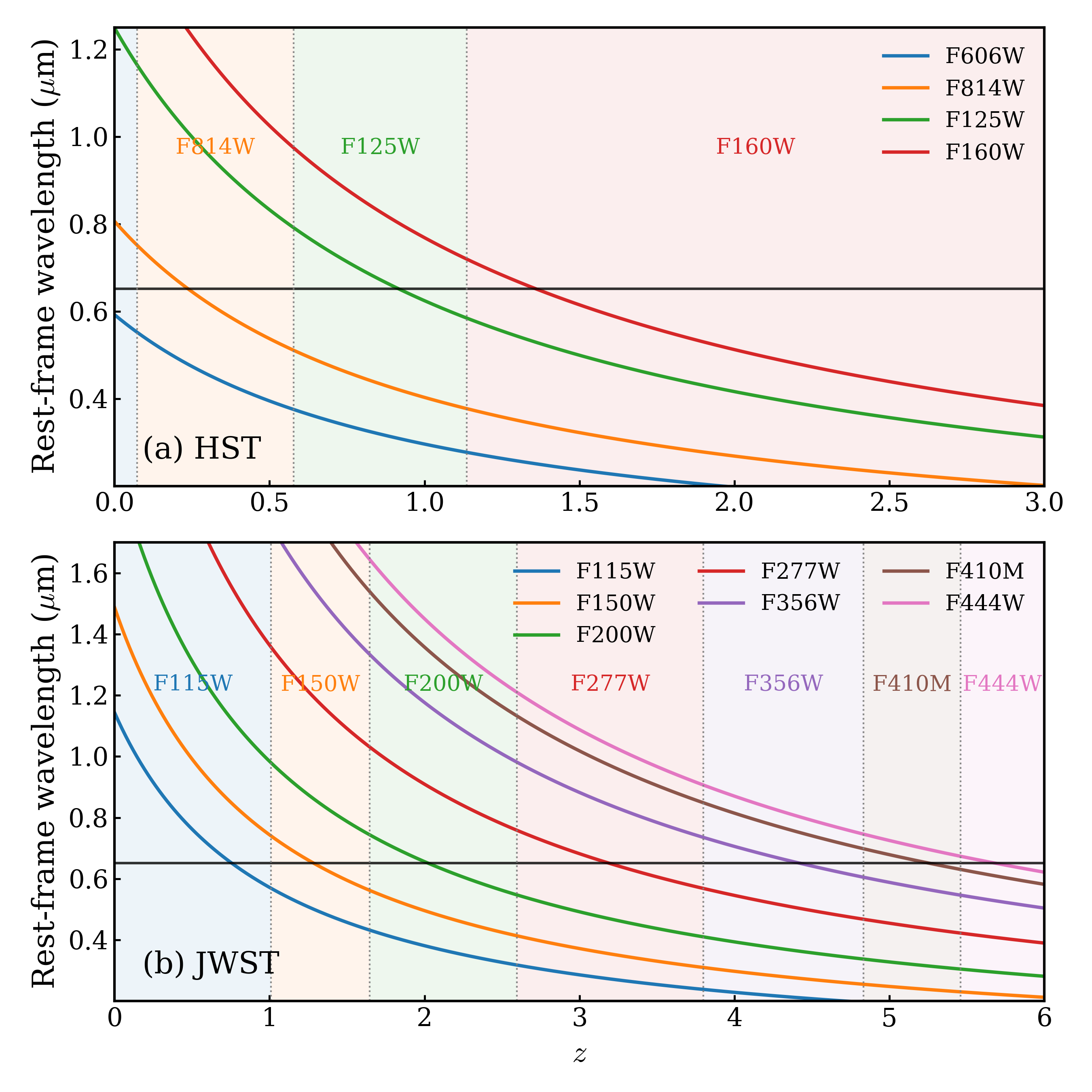}
\caption{Rest-frame wavelength of the filters versus redshift. The solid horizontal black line marks the target rest-frame wavelength of 6511.6 \AA, corresponding to the CGS $R$ band. The colored curves show the rest-frame wavelengths of the (a) HST (F606W to F160W) and (b) JWST (F115W to F444W) filters at a given redshift [$ \lambda_{\text{obs}} / (1+z)$]. Vertical dashed lines and shaded regions indicate the redshift ranges where a specific filter is selected as the best match for the target wavelength.}
\label{fig:combined_filters}
\end{figure}

To simulate the galaxy at a different redshift $z_f$, we calculate the expected flux change accounting for the evolution of the luminosity distance and the bandwidth-stretching effect. To minimize the dependence on the galaxy's specific spectral energy distribution (SED), we adopt a ``matched filter" strategy. As illustrated in Figure~\ref{fig:combined_filters}, for any given redshift, we select the observing HST or JWST filter whose rest-frame wavelength is closest to the CGS $R$ band ($6511.6$~\AA). For HST, we consider the Advanced Camera for Surveys (ACS) filters F606W and F814W and the Wide Field Camera 3 infrared-channel (WFC3/IR) filters F125W and F160W, and for JWST we choose the NIRCam filters (F115W, F150W, F200W, F277W, F356W, F410M, F444W). Figure~\ref{fig:combined_filters} shows that while the discrete nature of the filters prevents a perfect match, the selected filters (indicated by the shaded regions) consistently track the target wavelength. Assuming that the intrinsic specific luminosity $L_\nu$ varies mildly between the target wavelength and the effective rest-frame wavelength of the selected filter, we can neglect the residual color term. Thus, the flux scaling is governed primarily by the geometry and the bandwidth factor,

\begin{equation}
\frac{f_f}{f_i} \approx \left(\frac{D_{L,i}}{D_{L,f}}\right)^2 \left(\frac{1+z_f}{1+z_i}\right).
\end{equation}

\noindent
The corresponding magnitude transformation is then

\begin{align}
m_f - m_i &= (-2.5 \log f_f + M_{0,f}) - (-2.5 \log f_i + M_{0,i}) \notag \\
          &\approx -2.5 \log \left(\frac{D_{L,i}}{D_{L,f}}\right)^2 - 2.5 \log \left(\frac{1+z_f}{1+z_i}\right) \notag \\
          &\quad + (M_{0,f} - M_{0,i}).
\end{align}

\noindent
Here, the term $-2.5 \log [(1+z_f)/(1+z_i)]$ represents the bandwidth dimming effect, which must be included even when sampling similar rest-frame wavelengths. To mimic the realistic conditions of observations, we incorporate two types of noise into the mock images: Poisson noise associated with the source electron count and background noise related to the instrumentation and exposure time, which is mimicked using the real background of the target field. Noise was introduced by converting the image counts to electrons and adding random values based on Poisson and background noise distributions.

To account for the evolutionary trends for our galaxy sample, we derive empirical relations for luminosity and size evolution directly from the catalogs of the CANDELS EGS field (\citealt{grogin_candels_2011,koekemoer_candels_2011,stefanon_candels_2017}). We first select a sample of 10,043 star-forming galaxies as a statistically practical proxy for a disk-dominated population, because uniform visual disk classifications are not available over the full redshift range and the $UVJ$-quiescent population contains a larger fraction of spheroid-dominated systems. This proxy is not one-to-one: it may include irregular or non-disk galaxies and omit quiescent disks. The selected sample spans a redshift range of $0 \le z \le 6$ and a stellar mass range of $10^9$ to $10^{11.5}\, M_{\odot}$. The crucial step of separating star-forming galaxies from the quiescent population is performed using the well-established $UVJ$ color-selection technique, following the methodology of \citet{williams_detection_2009}. The final star-forming sample is then divided into five stellar mass bins, within each of which the galaxies are binned further into 10 equal-number redshift bins to trace their evolution. The sample size per redshift bin varies with stellar mass, ranging from $N = 574$ for the lowest mass bin ($10^{9.0}-10^{9.5}\, M_{\odot}$) to $N \approx 10$ for the most massive bin ($10^{11.0}-10^{11.5}\, M_{\odot}$), reflecting the natural scarcity of unquenched massive galaxies at high redshift in the limited survey volume. For each redshift bin, we compute the median rest-frame $R$-band absolute magnitude ($M_R$) and the median effective radius ($R_e$).
The effective radii were taken from the F160W-band measurements in the CANDELS EGS catalog and corrected to a rest-frame wavelength of approximately 6300~\AA, using the wavelength-dependent size correction of \citet{van_der_wel_3d-hstcandels_2014}. We adopt this wavelength to approximate the rest-frame $R$ band, rather than their reference wavelength of 5000~\AA. Finally, we fit parametric models to the binned data points for each mass bin independently to determine the evolutionary parameters. For luminosity evolution, we assume

\begin{equation}
M_R(z) = M_0 - 2.5 \alpha \log(1+z), 
\end{equation}

\noindent
and for size evolution we adopt the power-law model

\begin{equation}
R_e(z) = R_0 (1+z)^\beta.
\end{equation}

\noindent
The luminosity relation is equivalent to $L_R(z)/L_R(0)=(1+z)^\alpha$, wherein $\alpha=0$ denotes no luminosity evolution, $\alpha>0$ and $\alpha<0$ denote a brighter and fainter population at higher redshift, respectively, and $\alpha=1$ corresponds to luminosity increasing in direct proportion to $1+z$; larger $|\alpha|$ implies stronger evolution. Similarly, $\beta=0$ denotes no size evolution, $\beta<0$ and $\beta>0$ denote, respectively, smaller and larger galaxies at higher redshift, and a more negative $\beta$ implies stronger size growth toward low redshift.

\begin{figure}[ht!]
\includegraphics[width=\columnwidth]{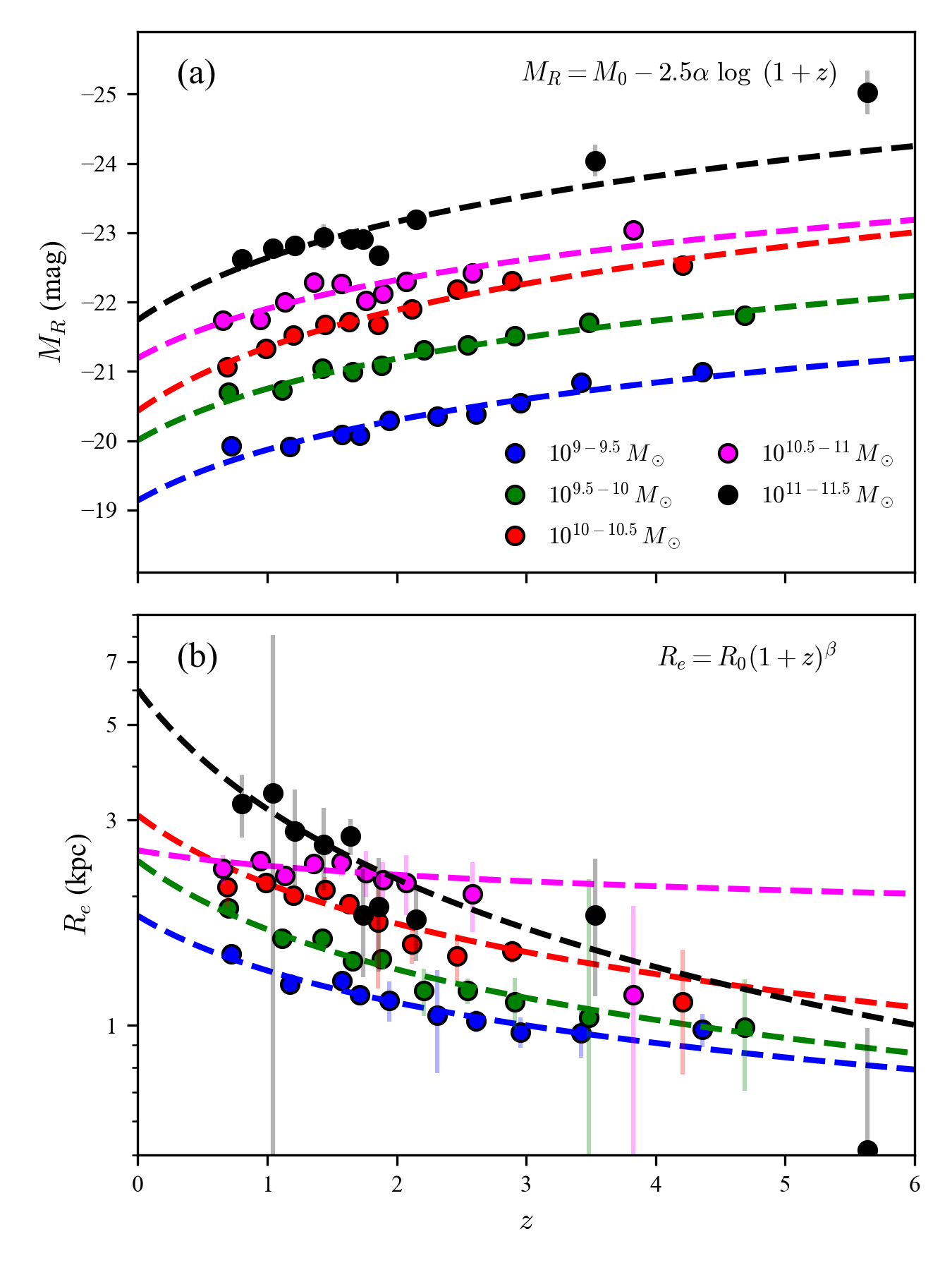}
\caption{Empirical relations for the (a) luminosity ($M_R$ vs. $z$) and (b) size ($R_e$ vs. $z$) evolution of star-forming galaxies in the EGS field, derived from CANDELS data. The data points represent the median values in redshift bins for five distinct stellar mass bins. Error bars show $\sigma/\sqrt{N}$, where $\sigma$ is the sample standard deviation and $N$ is the number of galaxies in the bin. Dashed lines show the best-fit parametric model for each mass bin.}
\label{fig:evolution_combined}
\end{figure}

These steps allow us to generate mock images that simulate the appearance of the selected barred spiral galaxies at various redshifts, incorporating realistic models of how both distance effects and intrinsic galaxy properties evolve over time. The results of this analysis are presented in Figure~\ref{fig:evolution_combined}, with the best-fit parameters for each mass bin summarized in Table~\ref{tab:fit_params}. The derived evolution parameters, $\alpha$ and $\beta$, are subsequently used to incorporate luminosity and size evolution effects into our mock image generation process.

\begin{deluxetable}{ccccc}[!htbp]
\tabletypesize{\footnotesize}
\tablecolumns{5}
\tablecaption{Best-fit Parameters for Luminosity and Size Evolution \label{tab:fit_params}}
\tablewidth{0pt}
\tablehead{
    \colhead{$\log M_*$} & \colhead{$M_0$} & \colhead{$\alpha$} & \colhead{$R_0$} & \colhead{$\beta$} \\
    \colhead{($M_\odot$)} & \colhead{(mag)} & \colhead{} & \colhead{(kpc)} & \colhead{} \\
    \colhead{(1)} & \colhead{(2)} & \colhead{(3)} & \colhead{(4)} & \colhead{(5)}
}
\startdata
9.0$-$9.5   & $-$19.14 & 0.97 & 1.80 & $-$0.42 \\
9.5$-$10.0  & $-$20.01 & 0.99 & 2.41 & $-$0.53 \\
10.0$-$10.5 & $-$20.43 & 1.22 & 3.08 & $-$0.53 \\
10.5$-$11.0 & $-$21.19 & 0.94 & 2.55 & $-$0.12 \\
11.0$-$11.5 & $-$21.74 & 1.19 & 6.02 & $-$0.92 \\
\enddata
\tablecomments{Col. (1): Stellar mass bin. Col. (2): Rest-frame $R$-band absolute magnitude at $z=0$. Col. (3): Dimensionless luminosity evolution parameter. Col. (4): Effective radius at $z=0$, corrected to a rest-frame wavelength of approximately 6300~\AA. Col. (5): Dimensionless size evolution parameter.}
\end{deluxetable}

\setlength{\tabcolsep}{2pt}
\begin{deluxetable*}{ccccccccrr}[!htbp]
\tabletypesize{\scriptsize}
\tablecolumns{10}
\tablecaption{Basic Properties of the Barred Disk Galaxy Sample \label{tab:cgs_sample_properties}}
\tablewidth{0pt}
\tablehead{
    \colhead{Name} & \colhead{$R_{\rm bar}$} & \colhead{$e_{\rm bar}$} & \colhead{$\Theta_{\rm bar}$} & \colhead{$(I_2/I_0)_{\rm bar}$} & \colhead{$\Phi_{2,\rm bar}$} & \colhead{$i$} & \colhead{$\log M_\ast$} & \colhead{Scale} & \colhead{$T$} \\
    \colhead{} & \colhead{(\arcsec)} & \colhead{} & \colhead{(\degr)} & \colhead{} & \colhead{(\degr)} & \colhead{(\degr)} & \colhead{($M_\odot$)} & \colhead{(kpc/$\arcmin$)} & \colhead{} \\
    \colhead{(1)} & \colhead{(2)} & \colhead{(3)} & \colhead{(4)} & \colhead{(5)} & \colhead{(6)} & \colhead{(7)} & \colhead{(8)} & \colhead{(9)} & \colhead{(10)}
}
\startdata
ESO\,186$-$G062 & $24.2 \pm 8.5$  & $0.66 \pm 0.01$ & $159.6 \pm 2.9$ & $0.45 \pm 0.03$ & $15.2 \pm 3.0$  & 28.4 & $9.78 \pm 0.03$  & 9.4  & $7.0$ \\
NGC\,1022      & $19.1 \pm 3.3$  & $0.46 \pm 0.01$ & $118 \pm 11$    & $0.42 \pm 0.03$ & $62 \pm 19$      & 24.7 & $10.05 \pm 0.08$ & 5.4  & $1.1$ \\
NGC\,1291      & $244.9 \pm 2.0$ & $0.41 \pm 0.01$ & $172.6 \pm 4.1$ & $0.59 \pm 0.01$ & $160.2 \pm 3.0$  & 26.3 & $10.68 \pm 0.20$ & 2.5  & $0.1$ \\
NGC\,1300      & $9 \pm 19$      & $0.74 \pm 0.02$ & $106.2 \pm 3.1$ & $0.76 \pm 0.01$ & $79 \pm 47$      & 25.4 & $10.58 \pm 0.20$ & 5.6  & $3.7$ \\
NGC\,1302      & $28.1 \pm 3.9$  & $0.32 \pm 0.01$ & $173.1 \pm 1.8$ & $0.22 \pm 0.03$ & $9 \pm 13$       & 21.3 & $10.42 \pm 0.08$ & 5.8  & $0.0$ \\
NGC\,1387      & $25.5 \pm 2.3$  & $0.27 \pm 0.01$ & $109.2 \pm 1.7$ & $0.36 \pm 0.03$ & $66.9 \pm 1.7$   & 23.3 & $10.68 \pm 0.10$ & 5.2  & $-2.9$ \\
NGC\,1533      & $24.3 \pm 3.7$  & $0.35 \pm 0.01$ & $171.5 \pm 1.4$ & $0.35 \pm 0.01$ & $2 \pm 11$       & 22.0 & $10.58 \pm 0.10$ & 5.4  & $-2.6$ \\
NGC\,1574      & $19.2 \pm 1.7$  & $0.27 \pm 0.01$ & $154 \pm 15$    & $0.28 \pm 0.03$ & $29.78 \pm 0.58$ & 15.2 & $10.75 \pm 0.11$ & 5.4  & $-2.9$ \\
NGC\,1640      & $31.3 \pm 3.4$  & $0.59 \pm 0.01$ & $47.2 \pm 2.1$  & $0.53 \pm 0.01$ & $138 \pm 63$    & 22.0 & $9.95 \pm 0.08$  & 5.6  & $2.9$ \\
NGC\,2217      & $39.4 \pm 3.8$  & $0.43 \pm 0.02$ & $117.8 \pm 1.6$ & $0.51 \pm 0.01$ & $53 \pm 40$     & 21.9 & $10.95 \pm 0.07$ & 5.7  & $-0.8$ \\
NGC\,3275      & $28.3 \pm 3.3$  & $0.51 \pm 0.01$ & $120 \pm 10$    & $0.46 \pm 0.01$ & $69 \pm 68$     & 25.4 & $10.98 \pm 0.11$ & 12.3 & $1.8$ \\
NGC\,3313      & $3.79 \pm 0.27$ & $0.28 \pm 0.01$ & $103.0 \pm 2.5$ & $0.43 \pm 0.03$ & $72.4 \pm 2.7$  & 23.4 & $10.59 \pm 0.20$ & 16.2 & $2.0$ \\
NGC\,3450      & $38.2 \pm 6.6$  & $0.62 \pm 0.02$ & $124.5 \pm 3.2$ & $0.41 \pm 0.04$ & $48.6 \pm 9.8$  & 27.6 & $11.15 \pm 0.11$ & 17.5 & $2.9$ \\
NGC\,3892      & $34.1 \pm 4.0$  & $0.44 \pm 0.01$ & $99.7 \pm 3.8$  & $0.40 \pm 0.01$ & $83 \pm 23$     & 18.9 & $10.50 \pm 0.09$ & 7.9  & $-1.0$ \\
NGC\,4902      & $26.1 \pm 4.1$  & $0.57 \pm 0.01$ & $75.9 \pm 3.7$  & $0.46 \pm 0.03$ & $112.6 \pm 3.9$ & 29.4 & $10.80 \pm 0.07$ & 11.4 & $2.9$ \\
NGC\,5101      & $50.1 \pm 5.9$  & $0.50 \pm 0.01$ & $122.6 \pm 1.6$ & $0.64 \pm 0.04$ & $59.9 \pm 1.9$  & 20.3 & $11.07 \pm 0.20$ & 8.0  & $0.1$ \\
NGC\,5135      & $34.0 \pm 4.7$  & $0.56 \pm 0.01$ & $122 \pm 11$    & $0.87 \pm 0.01$ & $64.0 \pm 7.5$  & 28.6 & $11.04 \pm 0.12$ & 17.5 & $2.0$ \\
NGC\,5156      & $13.9 \pm 1.8$  & $0.57 \pm 0.02$ & $119.7 \pm 6.6$ & $0.37 \pm 0.01$ & $40.3 \pm 9.5$  & 24.8 & $10.70 \pm 0.09$ & 11.5 & $4.1$ \\
NGC\,5643      & $56.5 \pm 5.4$  & $0.65 \pm 0.01$ & $86.3 \pm 3.3$  & $0.35 \pm 0.03$ & $100.3 \pm 2.8$ & 25.1 & $10.63 \pm 0.20$ & 4.9  & $4.9$ \\
NGC\,6782      & $26.6 \pm 2.3$  & $0.48 \pm 0.01$ & $2.1 \pm 1.1$   & $0.59 \pm 0.01$ & $18 \pm 18$     & 26.3 & $10.94 \pm 0.07$ & 15.4 & $0.7$ \\
NGC\,6814      & $18.1 \pm 2.4$  & $0.36 \pm 0.03$ & $26.8 \pm 9.0$  & $0.21 \pm 0.01$ & $160 \pm 62$    & 16.5 & $10.65 \pm 0.09$ & 6.6  & $3.9$ \\
NGC\,7424      & $14.2 \pm 4.2$  & $0.61 \pm 0.01$ & $128.6 \pm 3.5$ & $0.42 \pm 0.01$ & $51.8 \pm 4.6$  & 23.4 & $9.42 \pm 0.20$  & 3.4  & $5.9$ \\
NGC\,7552      & $48.8 \pm 8.4$  & $0.60 \pm 0.01$ & $103.1 \pm 7.3$ & $1.04 \pm 0.01$ & $70 \pm 99$     & 23.0 & $10.45 \pm 0.10$ & 5.0  & $2.2$ \\
\enddata
\tablecomments{Col. (1): Galaxy name. Col. (2): Bar semi-major axis length (intrinsic, deprojected; see \citealt{li_carnegie-irvine_2011}). Col. (3): Bar ellipticity. Col. (4): Bar position angle, east of north. Col. (5): Relative amplitude of the $m=2$ mode of the bar. Col. (6): Phase angle of the $m=2$ mode of the bar. Col. (7): Disk inclination angle, where $0\degr$ is face-on. Col. (8): Stellar mass and its uncertainty. Col. (9): Physical scale in kpc per arcminute. Col. (10): Numerical morphological type $T$, from \citet{ho_carnegie-irvine_2011}. The remaining parameters are from \citet{li_carnegie-irvine_2011}.}
\end{deluxetable*}
\setlength{\tabcolsep}{3pt}

\begin{deluxetable}{cccc}[!htbp]
\tabletypesize{\footnotesize}
\tablecolumns{4}
\tablecaption{Basic Properties of the Unbarred Disk Galaxy Control Sample \label{tab:unbarred_sample_properties}}
\tablewidth{0pt}
\tablehead{
    \colhead{Name} & \colhead{$i$} & \colhead{$\log M_\ast$} & \colhead{Scale} \\
    \colhead{} & \colhead{(\degr)} & \colhead{($M_\odot$)} & \colhead{(kpc/$\arcmin$)} \\
    \colhead{(1)} & \colhead{(2)} & \colhead{(3)} & \colhead{(4)}
}
\startdata
IC\,1993   & 16.9 & $9.84 \pm 0.15$  & 3.96  \\
IC\,2056   & 29.7 & $9.86 \pm 0.08$  & 5.97  \\
IC\,5325   & 29.8 & $10.13 \pm 0.09$ & 5.27  \\
IC\,5332   & 19.8 & $9.40 \pm 0.20$  & 2.44  \\
NGC\,245   & 27.2 & $10.40 \pm 0.07$ & 14.93 \\
NGC\,1309  & 24.4 & $10.01 \pm 0.05$ & 7.57  \\
NGC\,1317  & 26.5 & $10.42 \pm 0.12$ & 4.92  \\
NGC\,5324  & 24.9 & $10.41 \pm 0.06$ & 12.80 \\
NGC\,5468  & 24.4 & $10.18 \pm 0.20$ & 13.59 \\
NGC\,6215  & 21.1 & $10.16 \pm 0.04$ & 5.97  \\
NGC\,6699  & 18.6 & $10.72 \pm 0.08$ & 13.33 \\
NGC\,6935  & 27.2 & $11.06 \pm 0.12$ & 17.55 \\
NGC\,7213  & 20.7 & $11.06 \pm 0.08$ & 6.40  \\
NGC\,7371  & 18.9 & $10.19 \pm 0.07$ & 9.08  \\
NGC\,7727  & 22.2 & $10.64 \pm 0.08$ & 6.78  \\
\enddata
\tablecomments{Col. (1): Galaxy name. Col. (2): Disk inclination angle, where $0\degr$ is face-on. Col. (3): Stellar mass and its uncertainty. Col. (4): Physical scale in kpc per arcminute. Parameters are adopted from \citet{li_carnegie-irvine_2011}. }
\end{deluxetable}
 
\subsection{Mock Images of High-redshift Galaxies}
For specific science goals, we need to apply the above-mentioned mock procedure to the CGS sample selected based on the target sample, producing mock images of the galaxies placed at higher redshifts as observed by specific telescopes (e.g., HST for low to moderate redshifts and JWST for high redshifts). This ensures that the intrinsic properties of the mock sample match those of the target sample, accounting for luminosity and size evolution. To achieve this, we select the HST/JWST filters of the mock images based on their proximity to the CGS $R$-band rest-frame wavelength ($6511.6$\,\AA). The selection process is illustrated in Figure~\ref{fig:combined_filters}, showing the corresponding filters for HST (panel~a) and JWST (panel~b) as a function of redshift. This selection directly impacts the pixel scale and PSF properties used for the mock observations. Table~\ref{tab:filter_properties} summarizes the pixel scale and PSF FWHM for each filter used in this study.

The EGS is a well-studied region of the sky. HST observations in the EGS field were conducted as part of CANDELS, which aimed to document the first third of galactic evolution from $z = 8$ to 1.5 via deep imaging of more than 250,000 galaxies using ACS optical and WFC3/IR imaging \citep{grogin_candels_2011, koekemoer_candels_2011}. The CANDELS program includes filters optimized for lower redshift observations, such as F606W, F814W, F125W, and F160W. JWST extends this coverage with the CEERS program \citep{finkelstein_ceers_2023}, which provides deeper NIRCam imaging. CEERS focuses on investigating galaxies in the first 500 Myr after the Big Bang, leveraging JWST's unprecedented sensitivity and resolution to push the boundaries of galaxy formation studies. CEERS covers $\sim 100\,{\rm arcmin}^2$ of the EGS field, in a region supported by a rich set of HST/CANDELS multi-wavelength data. Together, these surveys allow comprehensive mock observations for galaxies across a wide redshift range.

Taking the EGS field as the target, we select NGC\,3450 from the CGS sample to conduct mock observations. NGC\,3450 was chosen because its moderate inclination ($i=27.6\degr$), well-resolved bar, and prominent spiral arms make it a useful test case for following bar recovery and possible confusion with outer non-axisymmetric structure as the image quality degrades. It is not intended to represent the full range of bar morphologies in the sample. Figure~\ref{fig:3} illustrates a redshift-aligned visual comparison for the HST CANDELS and JWST CEERS mock images at $z=0.5, 1.0, 2.0, 3.0,$ and $4.0$. As expected, morphological features fade with increasing redshift due to surface brightness dimming and resolution limits, but the degradation rates differ significantly between the two surveys. Galactic structure blurs rapidly in the HST mocks: the spiral arms become faint at $z=1.0$ and are barely discernible by $z=2$. By $z=3$, the bar and bulge are virtually indistinguishable, merging into a compact, featureless component.  In contrast, the JWST mocks demonstrate superior capability in resolving structures at higher redshifts. The bar and spiral arms remain clearly visible at $z=2$. Even at $z=3$, where the HST image shows no substructure, the JWST mock still reveals traces of the barred morphology, eventually becoming a compact source only at $z=4$.

\begin{deluxetable}{ccccc}[!htbp]
\tabletypesize{\footnotesize}
\tablecolumns{5}
\tablecaption{Pixel Scale and PSF FWHM of Selected Filters \label{tab:filter_properties}}
\tablewidth{0pt}
\tablehead{
    \colhead{Telescope} & \colhead{Filter} & \colhead{Wavelength} & \colhead{Pixel Scale} & \colhead{PSF FWHM} \\
    \colhead{} & \colhead{} & \colhead{(\AA)} & \colhead{(\arcsec)} & \colhead{(\arcsec)} \\
    \colhead{(1)} & \colhead{(2)} & \colhead{(3)} & \colhead{(4)} & \colhead{(5)}
}
\startdata
HST  & F606W &  5922 & 0.06 & 0.13  \\
HST  & F814W &  8059 & 0.06 & 0.11  \\
HST  & F125W & 12486 & 0.06 & 0.18  \\
HST  & F160W & 15370 & 0.06 & 0.19  \\
JWST & F115W & 11434 & 0.03 & 0.066 \\
JWST & F150W & 14873 & 0.03 & 0.070 \\
JWST & F200W & 19680 & 0.03 & 0.077 \\
JWST & F277W & 27279 & 0.03 & 0.123 \\
JWST & F356W & 35287 & 0.03 & 0.142 \\
JWST & F410M & 40723 & 0.03 & 0.155 \\
JWST & F444W & 43504 & 0.03 & 0.161 \\
\enddata
\tablecomments{Col. (1): Telescope name. Col. (2): Filter name. Col. (3): Central wavelength of the filter. Col. (4): Pixel scale. Col. (5): FWHM of the PSF.}
\end{deluxetable}

\section{Bar Identification and Measurement Methods}
\label{sec:methods}

This section describes the one-dimensional (1D) isophotal analysis used by our automated pipeline: ellipse fitting and Fourier analysis. The limitations of 2D multi-component decomposition are discussed in Section~\ref{subsec:galfit_limitations}. These methods collectively enable a detailed characterization of bar structures and the quantification of their properties across different redshifts.

\subsection{Isophotal Analysis Method}
\label{subsec:isophote_analysis}

Isophotal analysis forms the foundational data extraction step for our 1D bar identification pipeline. For each galaxy image, we derive radial profiles of key structural parameters using two complementary methods: ellipse fitting, which provides geometric information, and Fourier analysis, which quantifies non-axisymmetric features.

The ellipse fitting method uses the \texttt{Ellipse} class from the \texttt{Photutils} package \citep{bradley_astropyphotutils_2022} to fit a series of concentric, elliptical isophotes to the galaxy image. The process begins by subtracting the background sky level, which we estimate from the mean flux in the outer regions of the image, following a procedure similar to that described in \citet{stone_autoprof_2021}. We employ a two-step fitting strategy to ensure robust centering: an initial fit is performed where all ellipse parameters (center, ellipticity $e$, and position angle $\Theta$) are allowed to vary freely. The resulting galaxy center from this preliminary fit is then held fixed for a second, more constrained fitting run, from which we extract the final profiles. This procedure yields radial profiles of the surface brightness ($\mu$), $e$, and $\Theta$ as a function of the semi-major axis, which for notational simplicity we denote as $R$.

For the Fourier analysis, we quantify the strength and orientation of non-axisymmetric structures, such as bars, by decomposing the light distribution along each isophote. To perform this analysis, we first extract the outermost five elliptical isophotes from the ellipse fitting results. Then we calculate the median values of $e$ and $\Theta$ from these isophotes to define a series of elliptical rings with consistent geometry. For each specific $R$, we consider the elliptical ring that covers the pixels within its boundaries. To ensure that the signal is uniformly sampled in azimuth, we implement binned sampling in which each elliptical annulus is divided into a set of azimuthal bins. The mean intensity of all pixels falling within each bin is calculated, creating a uniformly sampled 1D intensity sequence. For the discrete Fourier transform, we index the $N$ angular bins by $\chi_j=2\pi j/N$ and write

\begin{equation}
\begin{split}
I(R,\chi) = a_0(R)
+ \sum_{m=1}^{\infty}
\bigl[
&a_m(R)\cos(m\chi) \\
&+ b_m(R)\sin(m\chi)
\bigr].
\end{split}
\label{eq:fourier_series}
\end{equation}

\noindent
where $a_0(R)$ represents the average intensity and the coefficients $a_m(R)$ and $b_m(R)$ describe the amplitude and phase of different modes. For bar analysis, we focus on the bisymmetric ($m=2$) mode. The relative amplitude of the bar perturbation, defined as $I_2/I_0$ with $I_2(R) = \sqrt{a_2(R)^2 + b_2(R)^2}$ and $I_0(R) = a_0(R)$, serves as our primary measure of bar strength. Unlike ellipticity, this normalization is not bounded by unity. We define the harmonic phase of the $m=2$ mode as
$\phi_2(R)=\arg\!\left[a_2(R)+\mathrm{i}\,b_2(R)\right]$,
with $\phi_2$ defined modulo $2\pi$. This is the phase quantity used in the Fourier bar-detection criterion. Because $\phi_2$ is the phase of a bisymmetric mode, it is not itself the geometric position angle of the structure.

The physical azimuths of the bin centers are
$\varphi_j=-\pi+(j+1/2)2\pi/N$, for $j=0,\ldots,N-1$.
With $x'$ along the outer-disk major axis and $y'$ perpendicular to it in the image plane, the corresponding elliptical coordinate is
$\varphi=\arg\!\left(x'/q+\mathrm{i}\,y'\right)$,
where $q$ is the outer-disk axis ratio. Accounting for the offset between the Fourier index coordinate $\chi_j$ and the physical bin-center azimuth, the geometric direction of the $m=2$ component is obtained from

\begin{align}
t_2 &=
\frac{1}{2}
\left(
\phi_2+\frac{2\pi}{N}
\right),
\\
\Theta_2 &=
\left[
\Theta_{\rm disk}
+
\frac{180^\circ}{\pi}
\arg\!\left(
q\cos t_2+\mathrm{i}\sin t_2
\right)
\right]
\nonumber\\
&\hspace{2.2em}
\bmod 180^\circ .
\label{eq:fourier_direction}
\end{align}

Here $t_2$ is in radians, and $\Theta_{\rm disk}$ is the outer-disk direction in the same image position-angle convention used for ellipse fitting. Thus, $\Theta_2$ can be compared directly with the ellipse position angle. For comparison with the Fourier angles tabulated by \citet{li_carnegie-irvine_2011}, we express the same geometric direction in their clockwise convention from the positive image $y$ axis as
$\Phi_2=(-\Theta_2)\bmod180^\circ$.
Both $\Theta_2$ and $\Phi_2$ are therefore defined modulo $180^\circ$. Formal uncertainties in the Fourier orientation are propagated from the uncertainty in $\phi_2$ while holding the outer-disk geometry fixed.

\subsection{Automated Bar Identification and Measurement}
\label{subsec:auto_detection}

Once the radial profiles are extracted, we employ a systematic, automated pipeline to identify and verify galactic bars. Our approach builds upon established frameworks \citep[e.g.,][]{menendez-delmestre_near-infrared_2007, aguerri_population_2009, li_carnegie-irvine_2011} but introduces a rigorous, statistically optimized selection strategy to adapt to redshift-dependent observational effects.

\subsubsection{Candidate Selection and Criteria Optimization}
\label{subsubsec:optimization}

The first step is to identify potential bar candidates by locating local maxima in the strength profiles ($e$ for the ellipse method, $I_2/I_0$ for the Fourier method). Candidates are ranked by the stability of their angular profiles ($\Theta$ or $\phi_2$) within the bar region, prioritizing structures with coherent orientation.

A critical aspect of our pipeline is the adaptive optimization of detection criteria to mitigate the effects of resolution degradation and surface brightness dimming at high redshift. We designed the optimization process around a unified objective function that evaluates the performance of both our analysis methods simultaneously.

For each analysis method (ellipse and Fourier), we define an individual performance score,

\begin{equation}
S_{\text{method}} = f_{\text{true}} \, (1 - f_{\text{false}}),
\end{equation}

\noindent
where $f_{\text{true}}$ is the bar recovery rate (sensitivity, defined as the fraction of known barred galaxies correctly identified) and $f_{\text{false}}$ is the false-detection rate relative to the control sample (defined as the fraction of galaxies classified as unbarred in CGS that are classified as barred by our pipeline). Using this definition, we calculate $S_{\text{ellipse}}$ and $S_{\text{Fourier}}$ for the ellipse and Fourier methods, respectively. To ensure the robustness of the entire pipeline, our global optimization objective is to maximize the average score of the two methods,

\begin{equation}
\langle S \rangle = \frac{S_{\text{ellipse}} + S_{\text{Fourier}}}{2}.
\end{equation}

\noindent
Maximizing $\langle S \rangle$ ensures that the selected criteria balance the strengths of both methods, avoiding parameter sets that might favor one method at the expense of the reliability of the other. We apply this objective function in two distinct calibration phases:

\begin{enumerate}
\item Local bar criteria ($z \approx 0$). We calibrate our detection thresholds using the sample of 23 barred galaxies and the control sample of 15 unbarred galaxies, both selected from CGS (Section~\ref{subsec:sample_selection}). These training sets are matched strictly in their range of stellar mass and inclination angle. Because the sample size is small, we can perform an exhaustive global grid search over the parameter space to identify the absolute maximum of $\langle S \rangle$. This optimization yields the following baseline parameters.

\begin{itemize}
\item Ellipse method: Peak $e > 0.38$, $\Theta$ stability $< 25.0\degr$, $e$ drop $> 0.1$, $\Theta$ change $> 10.0\degr$.
\item Fourier method: Peak $I_2/I_0 > 0.15$, $\phi_2$ stability $< 25.0\degr$, $I_2/I_0$ drop $> 0.1$.
\end{itemize}

\noindent
These values were chosen because they maximize the classification score in our local calibration, providing a baseline derived from high-quality observational experience at $z \approx 0$. As shown in Figure~\ref{fig:optimization}, the selected parameters (red circles) lie in the region of highest classification score (brightest regions), confirming their robustness.

\begin{figure*}
\centering
\includegraphics[width=\textwidth]{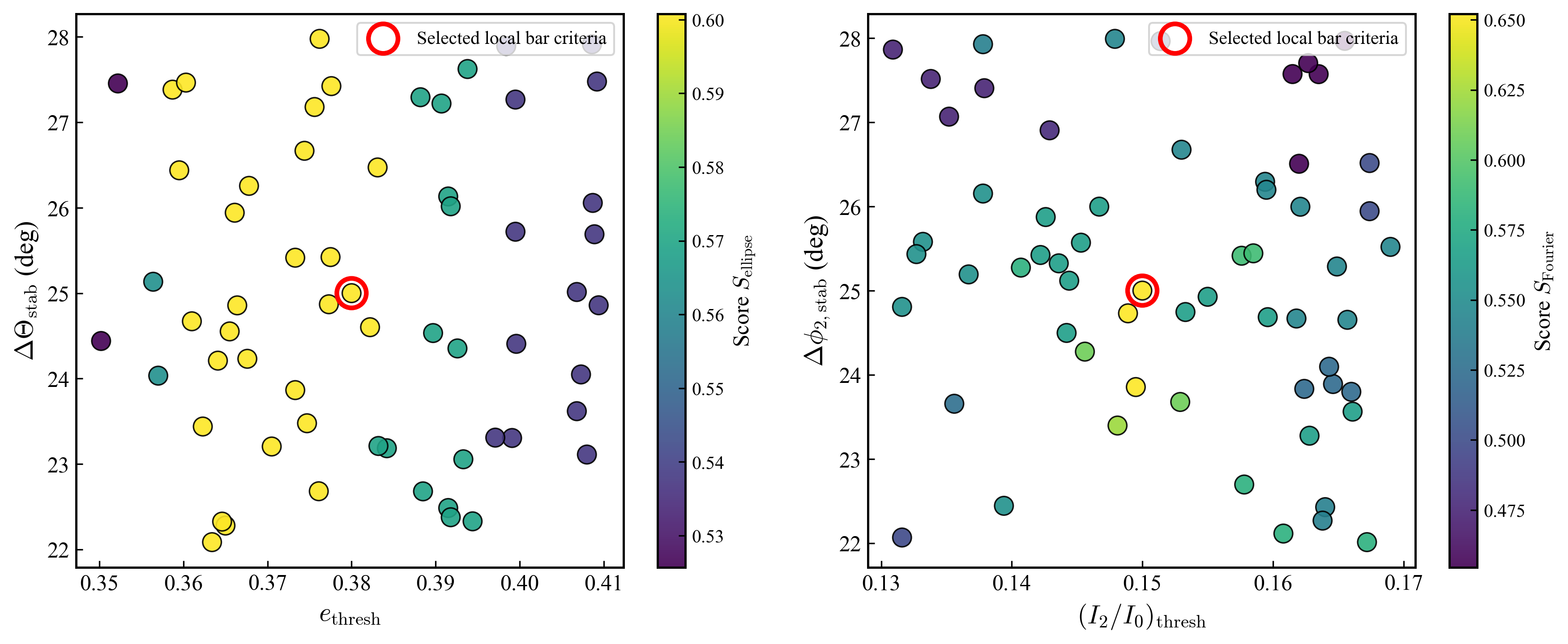}
\caption{Optimization of bar detection criteria for the local ($z \approx 0$) CGS sample. The color maps display the classification score $S = f_{\text{true}} (1 - f_{\text{false}})$ as a function of two key parameters for the ellipse (left) and Fourier (right) method. Brighter (yellow) regions indicate higher classification accuracy. The red circles mark the optimal parameter sets selected as our ``local bar criteria'': $(e_{\text{thresh}}, \Delta \Theta_{\text{stab}}) = (0.38, 25.0\degr)$ and $[(I_2/I_0)_{\text{thresh}}, \Delta \phi_{2,\text{stab}}] = (0.15, 25.0\degr)$. These values maximize the recovery of known bars while minimizing false detections relative to the unbarred control sample.}
\label{fig:optimization}
\end{figure*}

\item High-redshift bar criteria ($z > 0$). Applying local criteria directly to high-redshift images would introduce severe systematic biases. To address this, we derive optimal detection parameters specific to each survey (HST/CANDELS and JWST/CEERS) and redshift bin. Unlike the local calibration, which relies on observational experience, this approach leverages the known input classifications from our simulations: the barred sample defines the recovery rate, while the unbarred control sample provides a conservative contamination check. For each target redshift and survey configuration, we generate a specific training dataset by projecting the entire barred and unbarred sample (Section~\ref{subsec:sample_selection}) to the corresponding redshift using the methodology described in Section~\ref{sec:style}.
\end{enumerate}

We then seek the set of seven control parameters that maximizes the average score $\langle S \rangle$ for each specific redshift bin. Since the objective function is computationally expensive and potentially multimodal, we employ the efficient global optimization algorithm \citep{jones_efficient_1998}, which utilizes a Gaussian process (kriging) surrogate model to approximate the response surface of $\langle S \rangle$ and selects the next sampling point by maximizing the ``expected improvement'' criterion. This approach balances the exploitation of the approximating surface with the exploration of the parameter space, allowing us to efficiently converge on the optimal criteria for each epoch.

Table~\ref{tab:optimized_criteria} summarizes the specific optimized criteria derived for each survey and redshift bin. Alongside the parameters, we report both the bar recovery rate ($f_{\text{true}}$) and the false-detection rate ($f_{\text{false}}$) to fully characterize the performance. A clear trend is observed in the geometric parameters: the ellipticity threshold generally decreases from $e_{\text{thresh}} = 0.36$ at $z=0$ to $e_{\text{thresh}} \approx 0.03-0.27$ at high redshifts ($z \ge 1$), reflecting the practical reality that bars appear rounder because of PSF blurring. We acknowledge certain limitations in this automated optimization process. In a few specific bins (e.g., CEERS at $z=1.0$), the optimization may settle near boundary values. However, as shown by the scores in Table~\ref{tab:optimized_criteria}, the derived parameters consistently yield statistically reliable classifications across cosmic time, enabling a fully automated, uniform analysis.

{\setlength{\tabcolsep}{3pt}
\begin{deluxetable*}{lc|cccccc|ccccc|c}[!htbp]
\tabletypesize{\footnotesize}
\tablecolumns{14}
\tablecaption{Optimized High-redshift Bar Detection Criteria \label{tab:optimized_criteria}}
\tablewidth{0pt}
\tablehead{
    \colhead{Survey} & \colhead{$z$} & \multicolumn{6}{c}{Ellipse Method} & \multicolumn{5}{c}{Fourier Method} & \colhead{Overall} \\
    \colhead{Field}  & \colhead{}    & \multicolumn{4}{c}{Parameters} & \multicolumn{2}{c}{Performance} & \multicolumn{3}{c}{Parameters} & \multicolumn{2}{c}{Performance} & \colhead{Score} \\
    \colhead{}       & \colhead{}    & \colhead{$e_{\text{thresh}}$} & \colhead{$\Delta \Theta_{\text{stab}}$} & \colhead{$\Delta e_{\text{drop}}$} & \colhead{$\Delta \Theta_{\text{out}}$} & \colhead{$f^{\text{ellipse}}_{\text{true}}$} & \colhead{$f^{\text{ellipse}}_{\text{false}}$} & \colhead{$(I_2/I_0)_{\text{thresh}}$} & \colhead{$\Delta \phi_{2,\text{stab}}$} & \colhead{$\Delta(I_2/I_0)_{\text{drop}}$} & \colhead{$f^{\text{Fourier}}_{\text{true}}$} & \colhead{$f^{\text{Fourier}}_{\text{false}}$} & \colhead{$\langle S \rangle$} \\
    \colhead{}       & \colhead{}    & \colhead{} & \colhead{($^{\circ}$)} & \colhead{} & \colhead{($^{\circ}$)} & \colhead{} & \colhead{} & \colhead{} & \colhead{($^{\circ}$)} & \colhead{} & \colhead{} & \colhead{} & \colhead{} \\
    \colhead{(1)}    & \colhead{(2)} & \colhead{(3)} & \colhead{(4)} & \colhead{(5)} & \colhead{(6)} & \colhead{(7)} & \colhead{(8)} & \colhead{(9)} & \colhead{(10)} & \colhead{(11)} & \colhead{(12)} & \colhead{(13)} & \colhead{(14)}
}
\startdata
\textbf{CGS} (Ref) & 0.0 & 0.36 & 29.0 & 0.32 & 74.8 & 0.83 & 0.20 & 0.24 & 29.4 & 0.25 & 0.87 & 0.20 & 0.68 \\
\hline
\textbf{HST}       &     &      &      &      &      &      &      &      &      &      &      &      &      \\
EGS                & 0.5 & 0.27 & 13.4 & 0.20 & 49.2 & 0.70 & 0.07 & 0.11 & 43.6 & 0.19 & 0.82 & 0.33 & 0.60 \\
EGS                & 1.0 & 0.13 & 29.8 & 0.15 & 33.2 & 0.70 & 0.08 & 0.21 & 36.3 & 0.01 & 0.67 & 0.44 & 0.51 \\
EGS                & 2.0 & 0.04 & 12.0 & 0.04 & 63.3 & 0.76 & 0.23 & 0.03 & 44.3 & 0.07 & 0.65 & 0.39 & 0.49 \\
EGS                & 3.0 & 0.03 & 7.3  & 0.02 & 72.5 & 0.65 & 0.28 & 0.08 & 34.2 & 0.12 & 0.63 & 0.45 & 0.41 \\
\hline
\textbf{JWST}      &     &      &      &      &      &      &      &      &      &      &      &      &      \\
EGS                & 1.0 & 0.21 & 14.7 & 0.02 & 72.8 & 0.91 & 0.17 & 0.16 & 37.6 & 0.23 & 0.68 & 0.36 & 0.59 \\
EGS                & 2.0 & 0.23 & 18.8 & 0.23 & 68.3 & 0.73 & 0.07 & 0.13 & 27.6 & 0.17 & 0.76 & 0.49 & 0.53 \\
EGS                & 3.0 & 0.15 & 41.1 & 0.19 & 25.2 & 0.70 & 0.11 & 0.17 & 40.1 & 0.25 & 0.50 & 0.36 & 0.47 \\
EGS                & 4.0 & 0.21 & 27.0 & 0.17 & 8.6  & 0.70 & 0.11 & 0.27 & 38.8 & 0.14 & 0.47 & 0.37 & 0.46 \\
EGS                & 5.0 & 0.10 & 17.1 & 0.18 & 80.7 & 0.67 & 0.17 & 0.08 & 22.0 & 0.08 & 0.66 & 0.48 & 0.45 \\
EGS                & 6.0 & 0.21 & 8.8  & 0.13 & 1.0  & 0.71 & 0.15 & 0.12 & 39.5 & 0.19 & 0.57 & 0.39 & 0.48 \\
\enddata
\tablecomments{Col. (1): Target field name within the survey. Col. (2): Simulation redshift. Cols. (3)--(6): Optimized parameters for the ellipse method: $e_{\text{thresh}}$ (minimum peak ellipticity), $\Delta \Theta_{\text{stab}}$ (position angle stability threshold), $\Delta e_{\text{drop}}$ (ellipticity drop from peak), $\Delta \Theta_{\text{out}}$ (position angle change at outer boundary). Cols. (7)--(8): True positive ($f^{\text{ellipse}}_{\text{true}}$) and false-positive ($f^{\text{ellipse}}_{\text{false}}$) rates for the ellipse method. Cols. (9)--(11): Optimized parameters for the Fourier method: $(I_2/I_0)_{\text{thresh}}$ (minimum peak amplitude), $\Delta \phi_{2,\text{stab}}$ (phase angle stability), $\Delta(I_2/I_0)_{\text{drop}}$ (amplitude drop from peak). Cols. (12)--(13): True positive ($f^{\text{Fourier}}_{\text{true}}$) and false-positive ($f^{\text{Fourier}}_{\text{false}}$) rates for the Fourier method. Col. (14): The maximum average classification score achieved by the efficient global optimization algorithm.}
\end{deluxetable*}}
\subsubsection{Multi-step Verification Strategy}
\label{subsubsec:verification}

Each candidate galaxy undergoes a rigorous verification sequence designed to classify the structure and filter out false positives. Unlike previous studies that often rely on a single, rigid ellipticity threshold (e.g., $e_{\text{max}} > 0.4$), our approach implements a hierarchical classification logic. This allows us to recover nonstandard bar features---specifically those in galaxies with faint bulges or bars aligned with the disk major axis---that would otherwise be rejected by simpler criteria.

\begin{enumerate}
\item Inner boundary check. The algorithm searches for an inner boundary where the strength profile ($e$ or $I_2/I_0$) begins to rise (moving outward from the galaxy center). Most bars typically exhibit a dip in ellipticity near the center because of the presence of a bulge. If a clear rise is not found, the candidate is flagged as a potential ``weak bulge'' but is not discarded. This classification is crucial for late-type spirals where the bar structure may dominate the central potential, preventing the classic drop in ellipticity.

\item Stability check. The angular variation ($\Delta \Theta$ or $\Delta \phi_2$) within the bar region (from the inner boundary to the peak) must be smaller than the optimized stability threshold (e.g., $\Delta \Theta_{\text{stab}} < 25.0\degr$ for the local sample). This ensures that the identified structure is coherent and distinguishes bars from spiral arms, which exhibit continuous phase shifts. For the Fourier method, stability is evaluated as the maximum minus minimum of the unwrapped harmonic phase within this region. The $0^\circ$--$360^\circ$ representation in Figure~\ref{fig:method_illustration} is used only for display; it does not change this unwrapping or the detection thresholds.

\item Outer boundary check. We define the end of the bar by $\Delta e_{\text{drop}}$ or $\Delta(I_2/I_0)_{\text{drop}}$, where the bar strength drops significantly from the peak. For the ellipse method, we ideally expect a corresponding change in position angle as the isophotes transition from the bar to the disk. However, if the position angle change is insufficient ($< \Delta \Theta_{\text{out}}$) but the ellipticity drop is clear, the feature is classified as an ``aligned'' bar. This classification captures cases where the orientation of the bar is fortuitously close to the major axis of the outer disk, making the geometric transition subtle.
\end{enumerate}

For every ellipse-method detection, we require the ellipticity to decrease by at least $\Delta e_{\rm drop}$ after the candidate peak. A ``standard'' detection additionally requires a position angle change of at least $\Delta\Theta_{\rm out}$ at the outer boundary. If the ellipticity decrease is present but the position angle change is not, we classify the feature as an ``aligned'' bar. Thus, the aligned-bar relaxation applies only to the position angle criterion; it does not remove the requirement that the ellipticity profile rises to a peak and subsequently declines.

We also address the detectability of these substructures at high redshift. While ``weak bulge'' or ``aligned'' features are discernible in local high-resolution data, PSF blurring at high redshifts tends to circularize the central regions, causing ``weak bulge'' bars to mimic standard profiles or lead to non-detections. Our high-redshift bar criteria (Section~\ref{subsubsec:optimization}) partially mitigate this by adapting the thresholds, but the physical interpretation of these subclasses becomes naturally limited by resolution in the distant Universe.

\subsubsection{Bar Measurement}

For each verified bar, we extract three basic structural parameters corresponding to the peak of the strength profile ($e$ or $I_2/I_0$):

\begin{enumerate}
\item {\it Bar length}. For ellipse fitting, the projected bar length ($R_{\rm bar}$) is defined as the semi-major axis at the peak ellipticity. To account for projection effects, we calculate the deprojected intrinsic bar length, $R_{\rm bar}^{\rm depro}$, assuming that the outer disk is intrinsically circular. Following \citet{li_carnegie-irvine_2011},

\begin{equation}
R_{\rm bar}^{\rm depro} = R_{\rm bar} \sqrt{\cos^2 \Delta \Theta + \left( \frac{\sin \Delta \Theta }{1 - e_{\text{disk}}} \right)^2},
\end{equation}

\noindent
\noindent
where $\Delta\Theta$ is the difference between the mean outer-disk position angle and the mean bar position angle from the inner boundary to the ellipticity peak, with $e_{\text{disk}}$ the ellipticity and $\Theta_{\text{disk}}$ the position angle of the outer disk. For this correction, $e_{\text{disk}}$ and $\Theta_{\text{disk}}$ are taken as the arithmetic means over the outermost five fitted isophotes. The Fourier radii require no additional deprojection because the fixed-geometry elliptical annuli already use the outer-disk geometry.

\item {\it Bar strength}. We quantify the strength of the bar ($S_{\rm bar}$) using the peak value of the profile. For the ellipse method, this corresponds to the maximum ellipticity, denoted as $e_{\rm bar}$. For the Fourier method, it is the maximum relative amplitude of the bisymmetric mode, denoted as $(I_2/I_0)_{\rm bar}$.

\item {\it Bar orientation}. The orientation of the bar is defined by the angle at the radius of peak strength. We denote this as the position angle $\Theta_{\rm bar}$ for the ellipse method and the geometric direction $\Theta_{2,\rm bar}$ for the Fourier method. We use $\Phi_{2,\rm bar}$ only when expressing this direction in the convention of \citet{li_carnegie-irvine_2011}. The subscript $\mathrm{bar}$ denotes the value at the radius of peak strength.
\end{enumerate}

\subsubsection[An Example: Bar in NGC 3450]{An Example: Bar in NGC\,3450}
\label{subsubsec:example_ngc3450}

\begin{figure*}[ht!]
\centering
\begin{minipage}{0.32\textwidth}
\centering
\includegraphics[width=\linewidth]{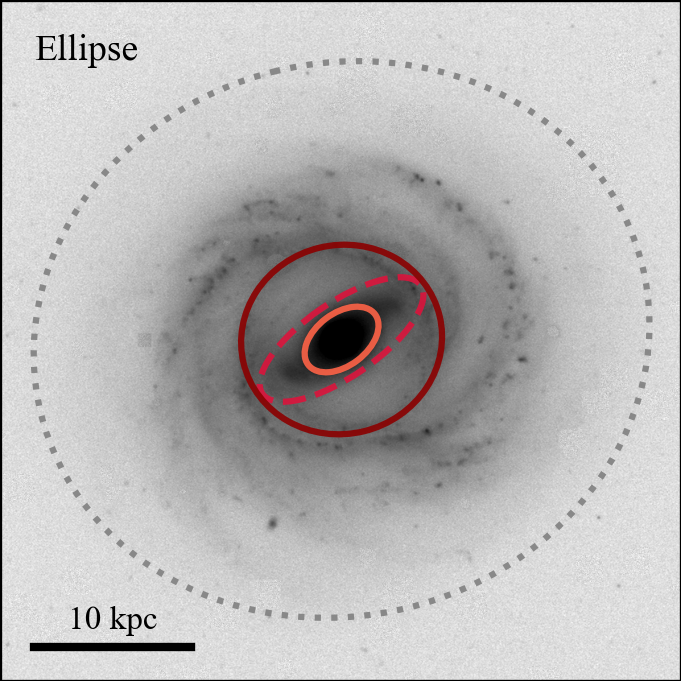}\vspace{1ex}

\includegraphics[width=\linewidth]{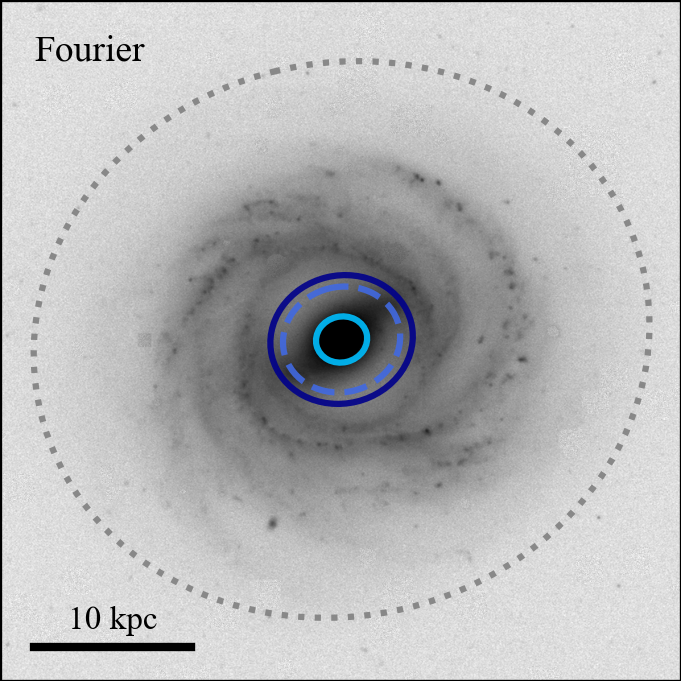}
\end{minipage}
\hfill
\begin{minipage}{0.64\textwidth}
\centering
\includegraphics[width=\linewidth]{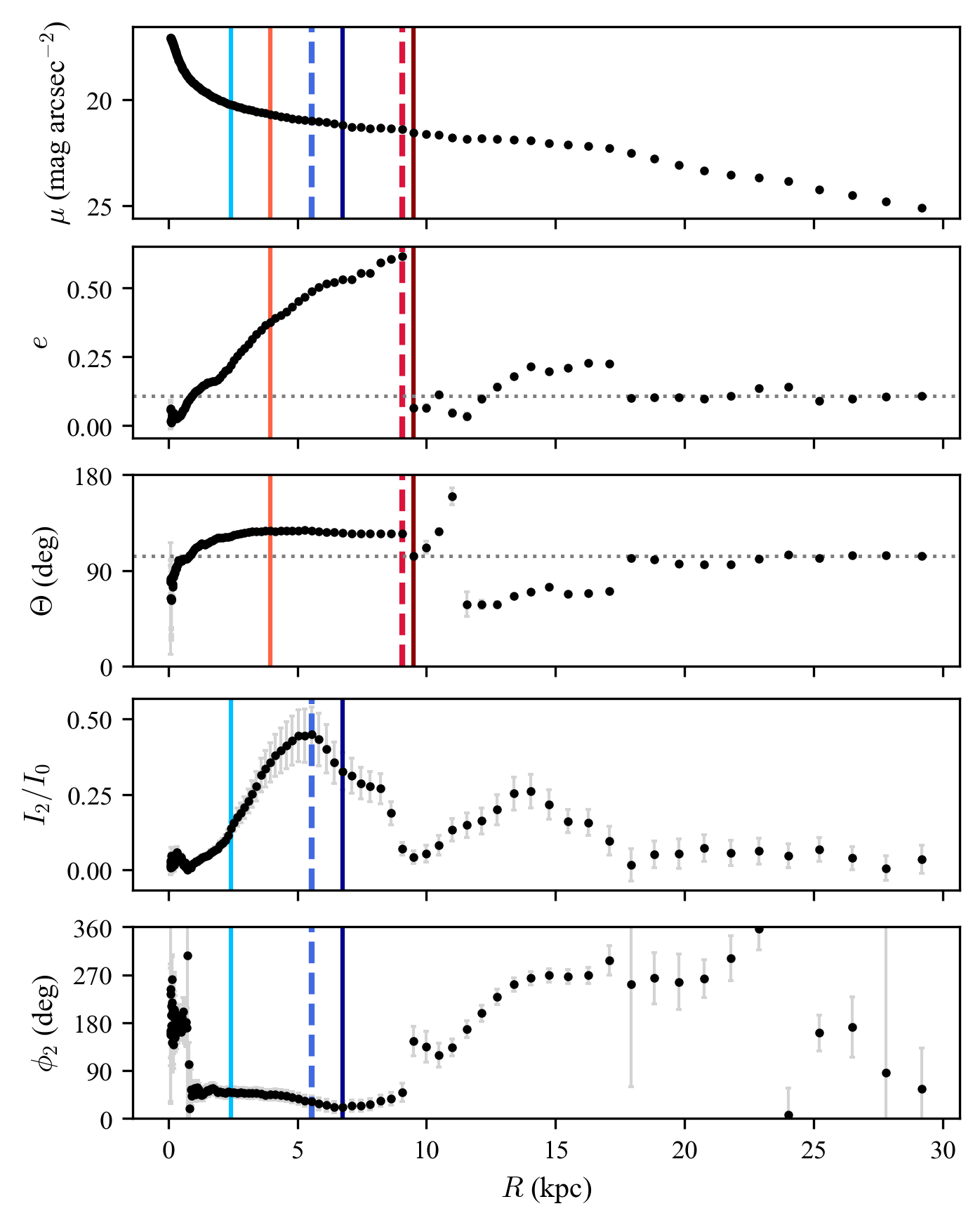}
\end{minipage}
\caption{An illustration of our automated bar identification and measurement pipeline applied to the CGS $R$-band image of NGC\,3450. The right panels give the five key radial profiles extracted from isophotal analysis plotted against the semi-major axis radius ($R$). From top to bottom, we plot the surface brightness ($\mu$), ellipticity ($e$), position angle ($\Theta$), Fourier amplitude ratio ($I_2/I_0$), and the harmonic phase of the second Fourier mode ($\phi_2$). Vertical lines indicate the bar parameters identified by our pipeline. For the ellipse method (red hues), the crimson dashed line marks the bar length ($R_{\rm bar}$) defined by the ellipticity peak, while tomato and dark red solid lines mark the inner and outer boundaries, respectively.  For the Fourier method (blue hues), the bar length and boundaries are marked by royal blue dashed and deep sky blue/dark blue solid lines.  The horizontal dotted lines in the $e$ and $\Theta$ profiles represent the values for the outer disk. The left panels give a visual inspection of the bar identified by the ellipse (top) and Fourier (bottom) method. The overlays follow the same color coding as the vertical lines in the right panels; the dashed ellipse corresponds to the measured bar length, while the solid ellipses mark the boundaries. The gray dotted ellipse represents the orientation and geometry of the outer disk used for reference. The two left images are displayed with an asinh stretch to reduce saturation in the central regions. Each image is scaled independently for display.}
\label{fig:method_illustration}
\end{figure*}

Figure~\ref{fig:method_illustration} illustrates our multi-step verification strategy applied to the CGS $R$-band image of NGC\,3450. The ellipse algorithm identifies a primary peak at a projected radius of $R_{\rm bar} = 31\farcs1$. Applying our local bar criteria (Section~\ref{subsubsec:optimization}), the peak strength of $e_{\rm bar} = 0.62$ significantly exceeds the optimized threshold of 0.38. A clear inner boundary is detected where the ellipticity rises moving outward, and the position angle remains stable with $\Delta \Theta \ll 25\degr$. The ellipticity decreases beyond the peak. This example is classified as an ``aligned'' bar: it satisfies the ellipticity-drop criterion while relaxing the outer position-angle-change requirement. After deprojection, the intrinsic bar length is determined to be $R_{\rm bar}^{\rm depro} = 31\farcs6$, with a position angle $\Theta_{\rm bar} = 124.4\degr$.

The Fourier method independently confirms this structure but yields slightly different quantitative results, a known characteristic arising from the different definitions of the two methods. It detects a well-defined peak in the $I_2/I_0$ profile with a strength of $(I_2/I_0)_{\rm bar} = 0.45$. The Fourier-derived bar length is $R_{\rm bar} = 19\farcs1$, shorter than the ellipse-based measurement, reflecting the fact that the $m=2$ Fourier amplitude typically peaks in the inner, stronger part of the bar, whereas the geometric ellipticity remains high further out. The ellipses marking the Fourier measurements follow the outer-disk geometry and identify radii, rather than the shape or orientation of the bar.

\section{Results}
\label{sec:results}

This section presents the quantitative results from our automated bar analysis pipeline. We first establish a robust local ($z \approx 0$) benchmark  using the high-quality CGS images. We then investigate how redshift-dependent observational effects, such as resolution degradation and surface brightness dimming, impact both the detectability and measurement accuracy of bars in our mock HST and JWST images. Throughout this analysis, we compare the performance of two distinct detection strategies: applying a fixed set of criteria derived from local galaxies (``local bar criteria'') versus applying criteria optimized for each specific redshift and survey (``high-redshift bar criteria'').

\subsection[Bar Properties at z approx. 0: The CGS Benchmark]{Bar Properties at $z \approx 0$: The CGS Benchmark}
\label{subsec:cgs_benchmark}

We first apply our pipeline to the 23 barred galaxies. The high-quality CGS images of these nearby galaxies provide an ideal testbed to validate the performance of our automated pipeline and establish a robust $z=0$ benchmark.

Table~\ref{tab:cgs_bar_fractions} summarizes the detection statistics. Our criteria, optimized as described in Section~\ref{subsubsec:optimization} to balance a high true positive rate with a low false-positive rate, are inherently stricter than those used in \citet{li_carnegie-irvine_2011}. Consequently, for this sample of previously identified barred galaxies, the ellipse method recovers 19 main bars ($f_{\rm bar} = 82.6\%$), whereas the Fourier method recovers 20 ($f_{\rm bar} = 87.0\%$). The ellipse method does not report a robust main-bar detection for NGC\,1291, NGC\,1302, NGC\,1387, and NGC\,1574, while the Fourier method misses the bar in NGC\,1300, NGC\,1302, and NGC\,6814. These non-detections do not imply that the galaxies are intrinsically unbarred. Instead, they indicate that the corresponding radial profiles do not satisfy our full set of deliberately strict main-bar criteria, including a robust signature of bar strength and well-defined bar boundaries.  Furthermore, our dual-candidate search successfully identifies internal bar-like substructures, or ``inner bars,'' in 7 of the 23 galaxies ($\sim 30\%$). A detailed analysis of these inner bar features is beyond the scope of this work but warrants future investigation.

\begin{figure*}[ht!]
\centering
\includegraphics[width=0.70\textwidth]{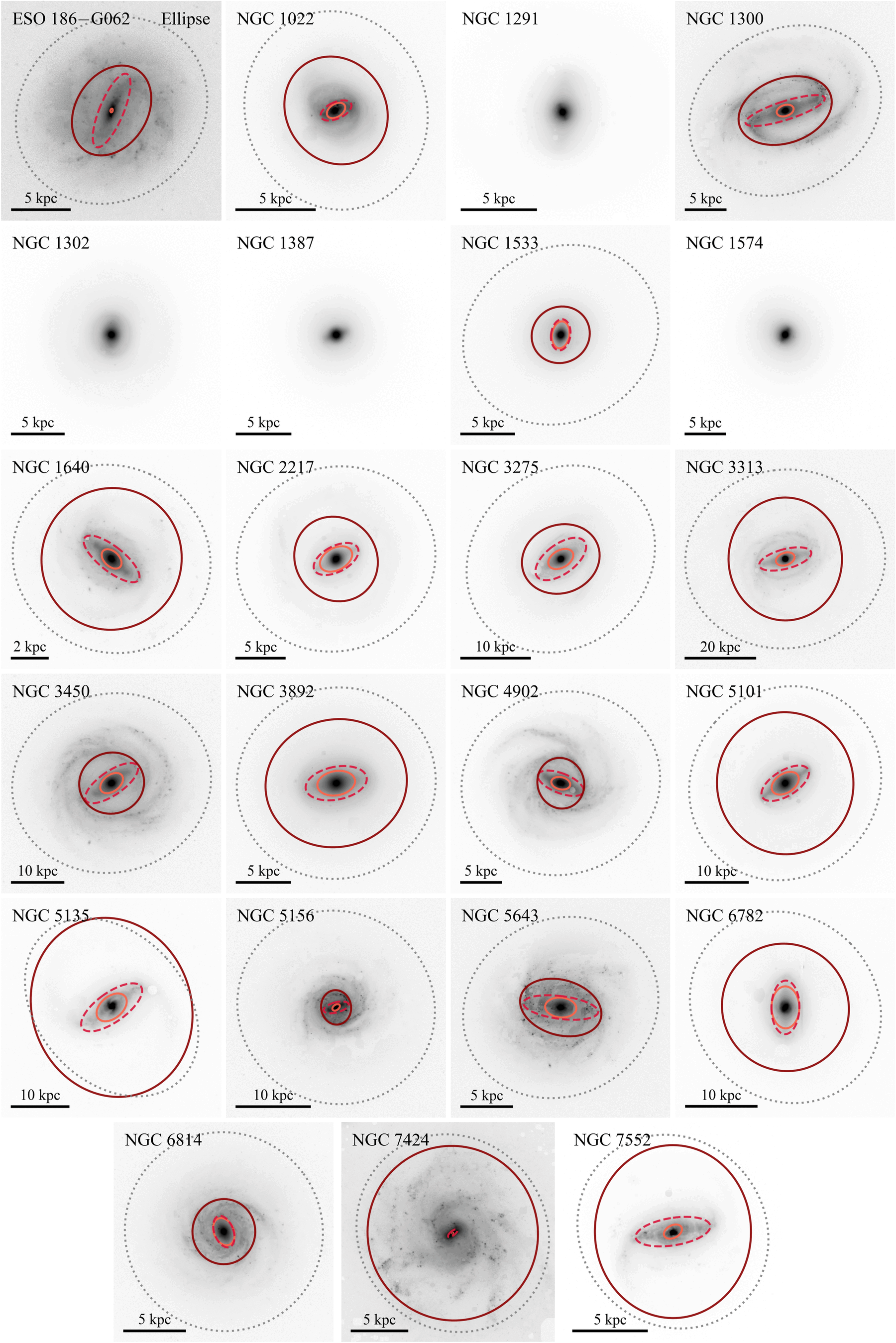}
\caption{Visual inspection of the CGS barred galaxy sample using the ellipse method. The overlays follow a color coding consistent with that defined in Figure~\ref{fig:method_illustration}: the crimson dashed ellipse corresponds to the measured bar length ($R_{\rm bar}$), while the tomato and dark red solid ellipses mark the inner and outer boundaries, respectively. The gray dotted ellipse represents the orientation and geometry of the outer disk. Panels without colored ellipses mark galaxies (NGC\,1291, NGC\,1302, NGC\,1387, and NGC\,1574) for which the ellipse method does not report a robust main-bar detection. The images are displayed with an asinh stretch to reduce saturation in the central regions. Each image is scaled independently for display and therefore does not provide a direct comparison of surface brightness between galaxies.}
\label{fig:cgs_mosaic_ellipse}
\end{figure*}

\begin{figure*}[ht!]
\centering
\includegraphics[width=0.70\textwidth]{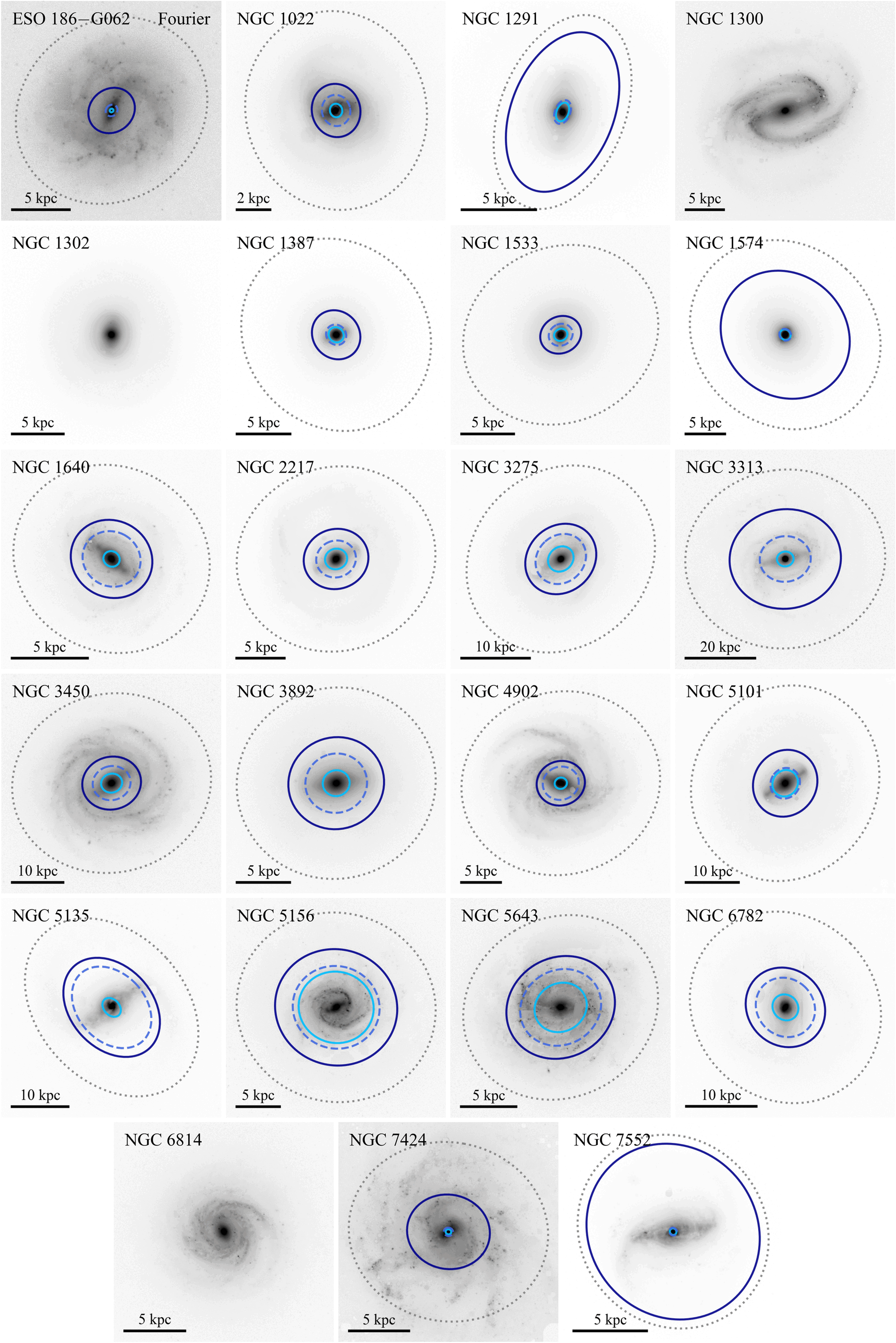}
\caption{Same as Figure~\ref{fig:cgs_mosaic_ellipse}, but for the Fourier method. The royal blue dashed ellipse corresponds to the measured bar length ($R_{\rm bar}$), defined by the peak of the $I_2/I_0$ profile, while the deep sky blue and dark blue solid ellipses mark the boundaries. While the method generally recovers the bar structure, in specific cases (e.g., NGC\,3313 and NGC\,5156) the peak $I_2/I_0$ is detected at a larger radius, likely influenced by spiral arms or outer rings that also exhibit significant bisymmetric ($m=2$) power. Panels without colored ellipses mark galaxies (NGC\,1300, NGC\,1302, and NGC\,6814) for which the Fourier method does not report a robust main-bar detection. The images are displayed with an asinh stretch to reduce saturation in the central regions. Each image is scaled independently for display and therefore does not provide a direct comparison of surface brightness between galaxies.}
\label{fig:cgs_mosaic_fourier}
\end{figure*}

To visually verify the performance of our automated pipeline, we present all 23 barred galaxies in Figures~\ref{fig:cgs_mosaic_ellipse} (ellipse method) and \ref{fig:cgs_mosaic_fourier} (Fourier method). Panels with colored bar overlays correspond to robust main-bar detections, while panels without such overlays indicate galaxies for which the corresponding method does not report a robust main-bar detection. For the ellipse method, the identified bar boundaries (solid red lines) and peak radii (dashed red lines) show excellent agreement with the visual morphology of the bars, confirming that our geometric criteria accurately isolate the bar component from the bulge and disk. The Fourier method also successfully recovers the vast majority of bars; however, this method occasionally identifies a peak $I_2/I_0$ amplitude at a radius larger than the visual bar end (e.g., in NGC\,3313 and NGC\,5156). This behavior is expected, as the Fourier method is sensitive to any non-axisymmetric $m=2$ mode, including spiral arms or rings that may persist beyond the bar itself.

\begin{figure*}[p]
\centering
\includegraphics[width=0.9\textwidth]{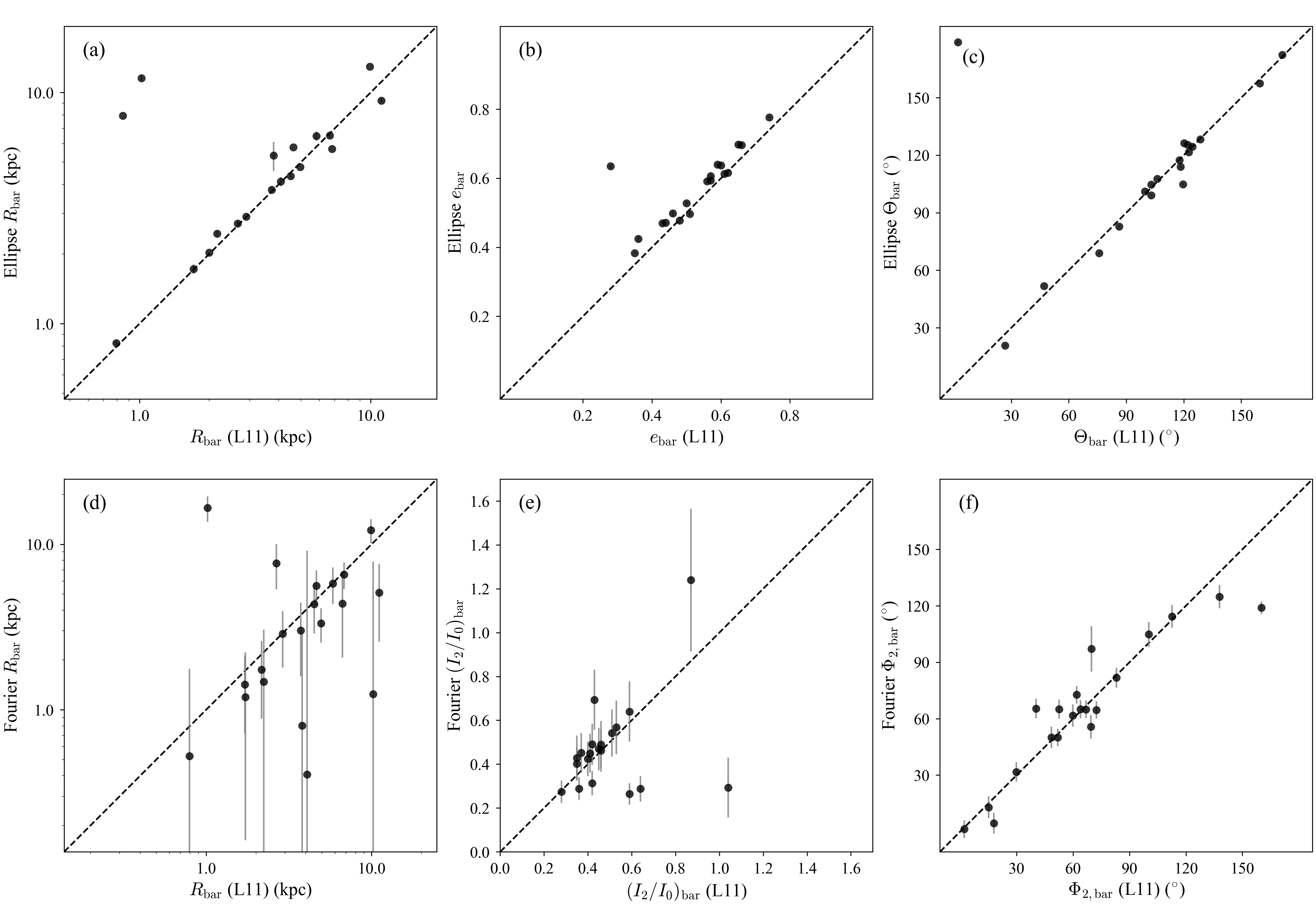}
\caption{Direct comparison of bar properties for the CGS sample measured by our automated pipeline versus those reported by \citet[][L11]{li_carnegie-irvine_2011}. The top row shows the results for the 19 galaxies successfully identified by the ellipse method for (a) intrinsic bar length ($R_{\rm bar}$), (b) maximum ellipticity ($e_{\rm bar}$), and (c) position angle ($\Theta_{\rm bar}$), while the bottom row gives the results for the 20 galaxies measured with the Fourier method for (d) $R_{\rm bar}$, (e) maximum Fourier amplitude [$(I_2/I_0)_{\rm bar}$], and (f) phase angle ($\Phi_{2, \rm bar}$). The non-detections for the ellipse method omitted from the top row are NGC\,1291, NGC\,1302, NGC\,1387, and NGC\,1574, while the non-detections for the Fourier method omitted from the bottom row are NGC\,1300, NGC\,1302, and NGC\,6814. In all panels, the black dashed line represents the 1:1 relation. }
\label{fig:cgs_li_comparison}
\end{figure*}

To further validate our automated measurements quantitatively, we compare our results directly with those from \citet{li_carnegie-irvine_2011}, as listed in Table~\ref{tab:cgs_sample_properties}. Figure~\ref{fig:cgs_li_comparison} presents a comprehensive object-by-object comparison. The bar length comparisons are shown on a logarithmic scale to better visualize the range of bar sizes and the consistency at smaller scales. We find good overall agreement between our measurements and the literature values, with data points clustering tightly around the 1:1 line. Minor discrepancies can be attributed to subtle differences in the fitting algorithms or specific criteria used to define the bar end. For the Fourier orientation comparison, we use $\Phi_{2,\rm bar}$ so that both axes follow the convention of \citet{li_carnegie-irvine_2011}.

\begin{figure*}[p]
\centering
\includegraphics[width=0.8\textwidth]{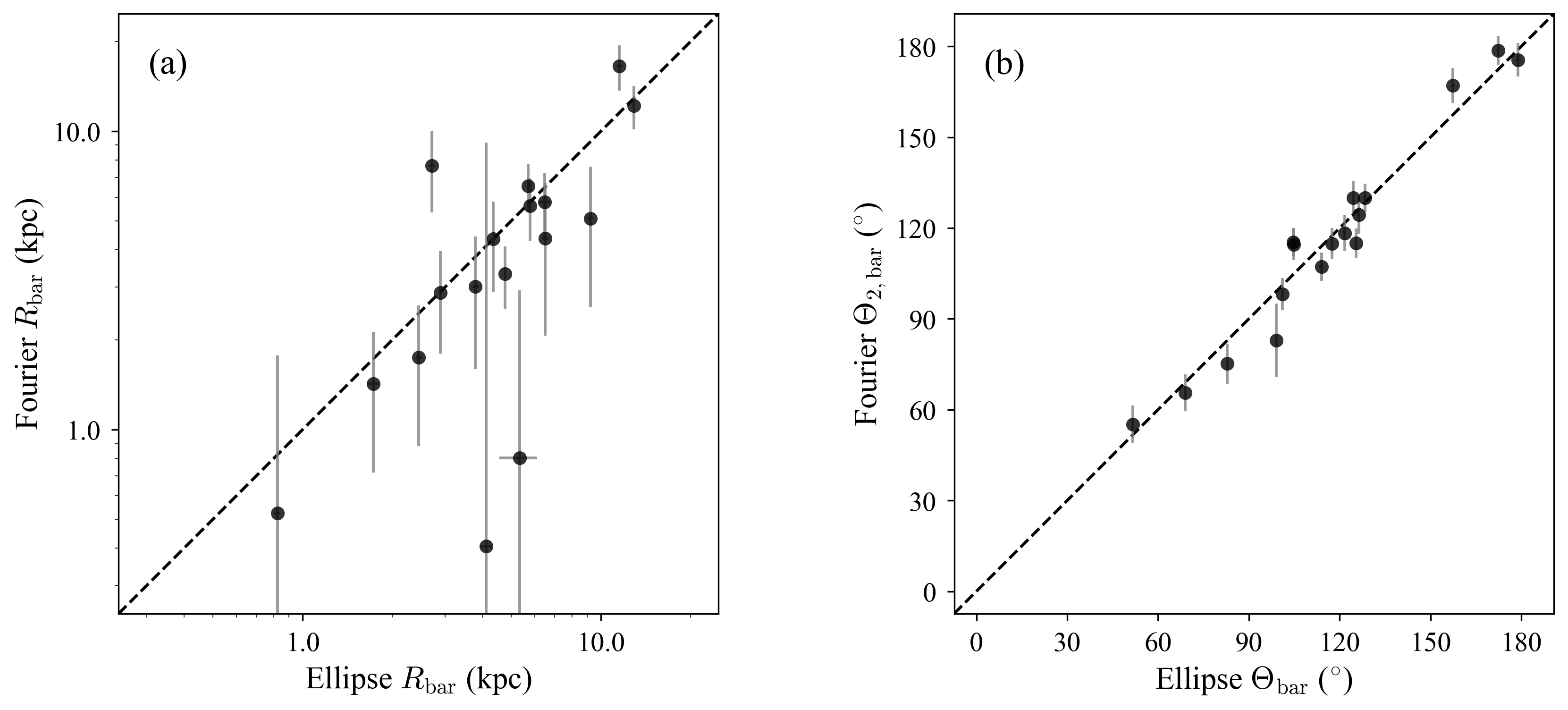}
\caption{Internal comparison of (a) intrinsic bar length ($R_{\rm bar}$) and (b) position angle ($\Theta_{\rm bar}$) versus the Fourier orientation ($\Theta_{2,\rm bar}$) for the 17 galaxies successfully measured by both the ellipse and Fourier methods.  The dashed line represents the 1:1 relation.}
\label{fig:cgs_method_comparison}
\end{figure*}

Finally, we directly compare the results from our two independent methods for the subset of 17 galaxies where both methods identified a main bar (Figure~\ref{fig:cgs_method_comparison}). A key finding emerges: the Fourier method systematically measures smaller bar radii than the ellipse method. This suggests that the peak of the $m=2$ Fourier amplitude profile is typically located at a smaller radius than the peak of the ellipticity profile. After converting the Fourier measurements to the same image position-angle convention as the ellipse measurements (Equation~\ref{eq:fourier_direction}), the bar orientations show general consistency.

\begin{deluxetable}{ccc}[!htbp]
\tabletypesize{\footnotesize}
\tablecolumns{3}
\tablecaption{Bar Detection Statistics for the CGS Sample \label{tab:cgs_bar_fractions}}
\tablewidth{0pt}
\tablehead{
    \colhead{Detection method} & \colhead{Count} & \colhead{Fraction} \\
    \colhead{} & \colhead{} & \colhead{(\%)} \\
    \colhead{(1)} & \colhead{(2)} & \colhead{(3)}
}
\startdata
Ellipse method              & 19 & 82.6 \\
Fourier method              & 20 & 87.0 \\
Successful in both methods  & 17 & 73.9 \\
Successful in either method & 22 & 95.7 \\
\enddata
\tablecomments{Col. (1): Bar detection method or combination of methods. Col. (2): Number of unique galaxies in the CGS sample that satisfy the detection criteria. Col. (3): Corresponding fraction of the total sample ($N=23$).}
\end{deluxetable}

\subsection{Impact of Redshift-dependent Observational Effects on Bar Detection}
\label{subsec:apparent_evolution}

Having validated our pipeline on the local CGS sample, we now turn to the primary goal of this paper: quantifying how redshift-dependent observational effects impact the detection of galactic bars. We applied our automated pipeline to the mock images simulating observations by HST and JWST. We carefully design the mock experiment to ensure sufficient statistical power. For each of the 23 barred galaxies at each target redshift, we generate 5 independent mock realizations by adding different real background noise maps cut from the target survey fields (CANDELS or CEERS). This procedure yields a total sample size of $23 \times 5 = 115$ mock observations at each redshift step. By treating these as distinct statistical trials, we can better estimate the variance in detectability introduced by background fluctuations and photon noise.

Figure~\ref{fig:bar_fraction_comparison} compares the bar recovery rates obtained using the static ``local bar criteria'' and the adaptive ``high-redshift bar criteria.'' We define the ``observed bar fraction'' ($f_{\text{obs}}$) as the percentage of the 115 mock galaxies in which a bar is successfully recovered by our pipeline. Since our input sample consists entirely of barred galaxies, an ideal pipeline would yield $f_{\text{obs}} = 100\%$. Thus, $f_{\text{obs}}$ is equivalent to the bar recovery rate for these simulations, while $1-f_{\text{obs}}$ gives the unrecovered fraction. Variations in $f_{\text{obs}}$ with redshift should not be interpreted as intrinsic cosmic evolution of the bar fraction, but instead as apparent evolution induced solely by resolution limits, surface brightness dimming, and band-shifting effects.

\subsubsection{Results with Local Bar Criteria}\label{subsubsec:quantifying_bias}
When applying the static criteria optimized for $z=0$ (Figure~\ref{fig:bar_fraction_comparison}a), we observe a significant decline in bar detectability with redshift, particularly for the ellipse method applied to HST data. For the HST/CANDELS simulations (blue circles), the recovery rate drops from $\sim 83\%$ at $z=0$ to $\sim 52\%$ at $z=0.5$, and further to $\sim 22\%$ by $z=1.0$. This implies that nearly 80\% of intrinsically barred galaxies would be classified as unbarred at $z \approx 1$ in HST imaging if local geometrical standards are applied rigidly. The Fourier method (orange circles) appears more robust in the HST regime, maintaining a nominally high recovery rate of $\sim 75\%$ at $z=1.0$. However, this higher detection rate must be interpreted with caution. Unlike the ellipse method, which strictly requires isophotal alignment, the Fourier method is sensitive to any global bisymmetric ($m=2$) signal. As illustrated later in Figure~\ref{fig:ngc3450_z_evolution}, top two rows, at high redshift in the HST/CANDELS mocks, the peak $m=2$ amplitude can shift to larger radii corresponding to spiral arms or outer rings rather than the bar itself. In the JWST/CEERS example, the same failure mode is less severe at $z=2.0$, but the Fourier measurement still becomes unreliable at higher redshift where the pipeline no longer reports a robust detection. Consequently, the apparent robustness of the Fourier method is partially driven by contamination from non-bar structures, a conclusion further supported by its consistently higher false-positive rates compared to the ellipse method (see Table~\ref{tab:optimized_criteria}). For the JWST/CEERS simulations (triangles), the higher resolution mitigates the decline for the ellipse method, but it still shows a steady drop to $\sim 35\%$ by $z=4.0$ if static criteria are used.

\subsubsection{Results with High-redshift Bar Criteria}
\begin{figure*}[ht!]
\centering
\begin{minipage}[b]{0.48\textwidth}
\centering
\includegraphics[width=\textwidth]{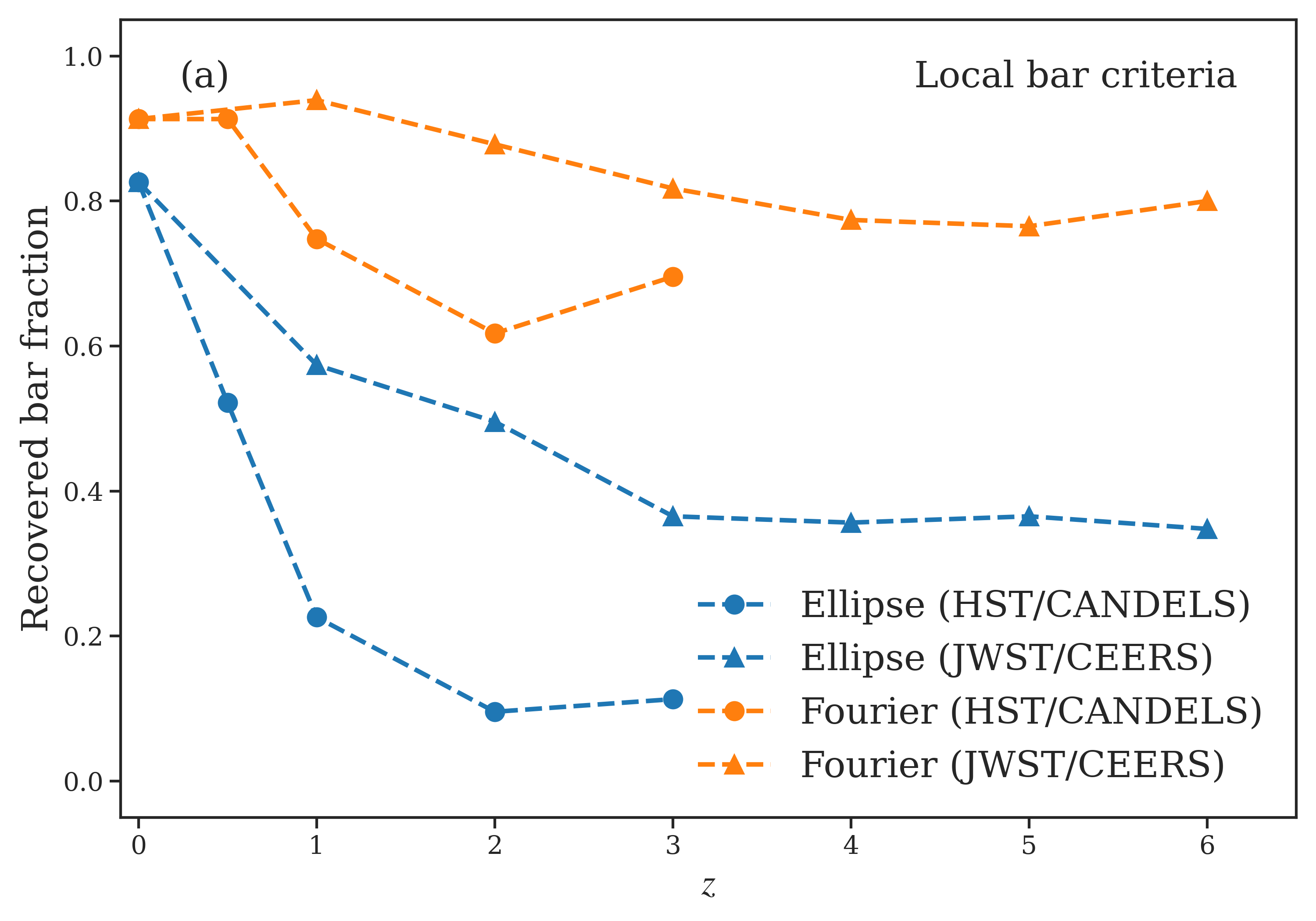}
\end{minipage}
\hfill
\begin{minipage}[b]{0.48\textwidth}
\centering
\includegraphics[width=\textwidth]{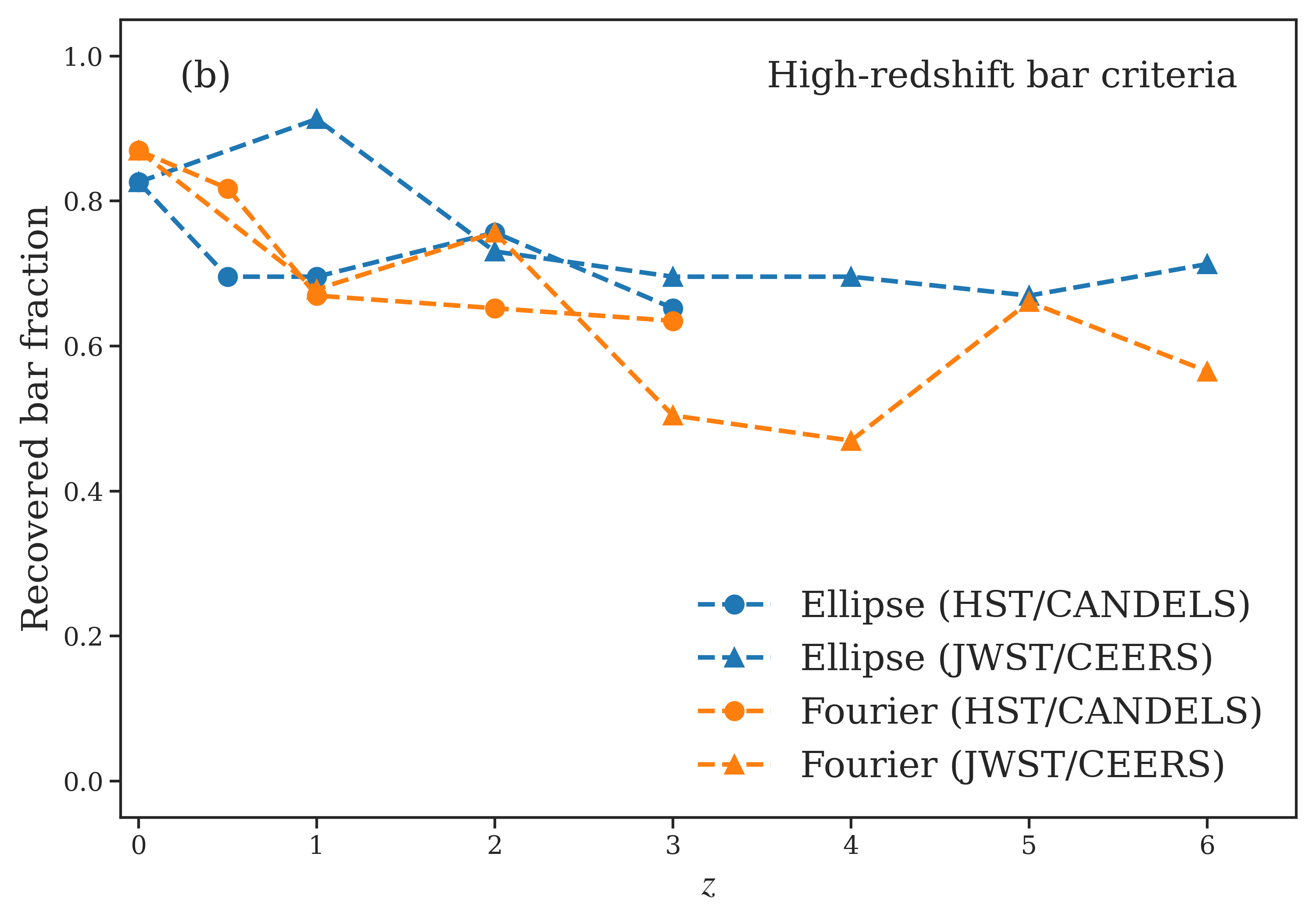}
\end{minipage}
\caption{The recovered bar fraction (recovery rate) as a function of redshift for a sample of intrinsically barred galaxies. The $y$-axis shows the fraction of successful detections out of 115 independent mock observations (23 galaxies $\times$ 5 backgrounds). Results are shown for mocks that use (a) static criteria optimized for local galaxies ($z \approx 0$) and (b) adaptive criteria optimized for each redshift (Table~\ref{tab:optimized_criteria}). Note that the local bar criteria yield an artificial, rapid decline in the bar fraction, especially for HST data, while the recovery rates using the high-redshift criteria are significantly higher, demonstrating the importance of adapting detection thresholds to observational conditions.}
\label{fig:bar_fraction_comparison}
\end{figure*}

Applying our adaptively optimized criteria (Figure~\ref{fig:bar_fraction_comparison}b) yields a dramatic recovery of lost bars. For the HST/CANDELS data, the ellipse method (blue circles) now maintains a remarkably high recovery rate, staying above $60\%$. Crucially, it sustains a rate of $\sim 65\%$ at $z=3.0$, a regime where the local criteria failed completely. This improvement is driven by the relaxed ellipticity thresholds (Table~\ref{tab:optimized_criteria}), which account for the physical rounding of bars by the PSF. For JWST/CEERS, the ellipse recovery rate remains above $60\%$ through $z=6$, while the Fourier recovery rate ranges from $47\%$ to $76\%$. This demonstrates that the apparent paucity of bars at high redshift reported in some studies may be largely alleviated by adapting detection thresholds to the observational reality of high-$z$ imaging.

\begin{figure*}[ht!]
\centering
\includegraphics[width=\textwidth]{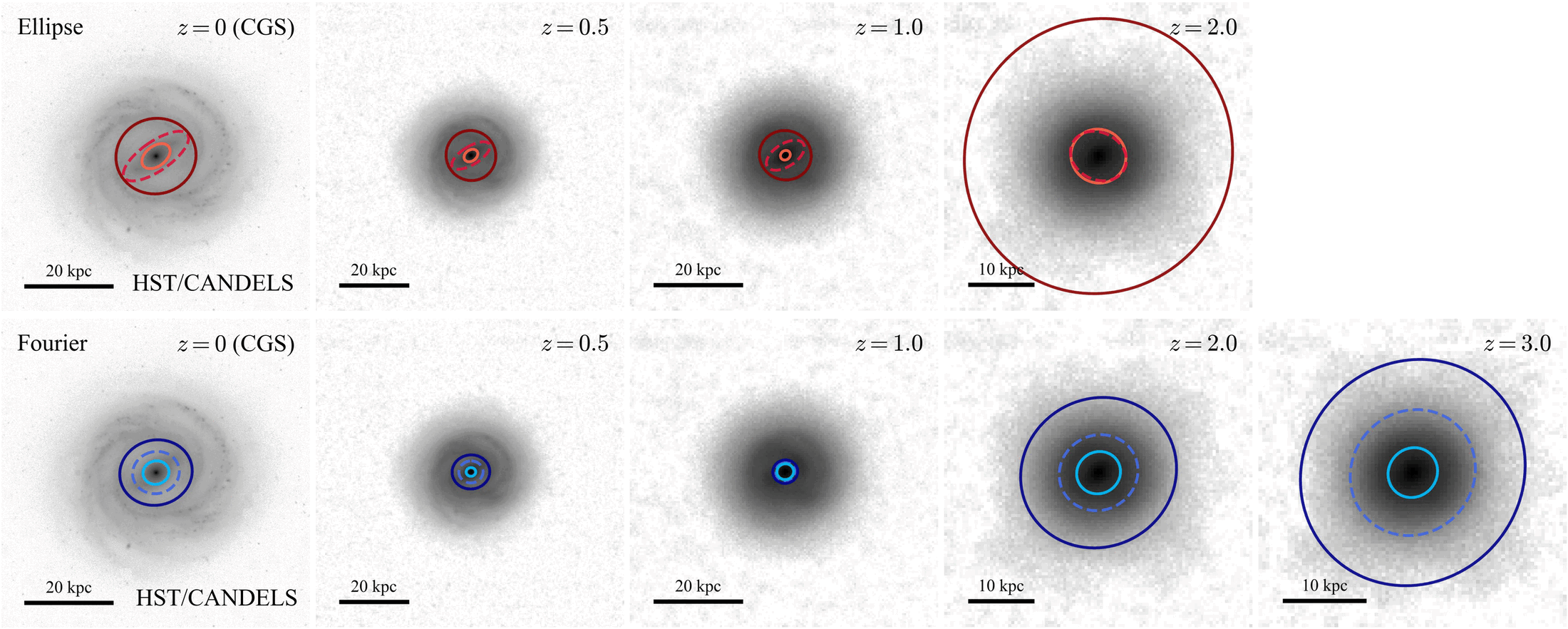}
\par\bigskip 
\includegraphics[width=\textwidth]{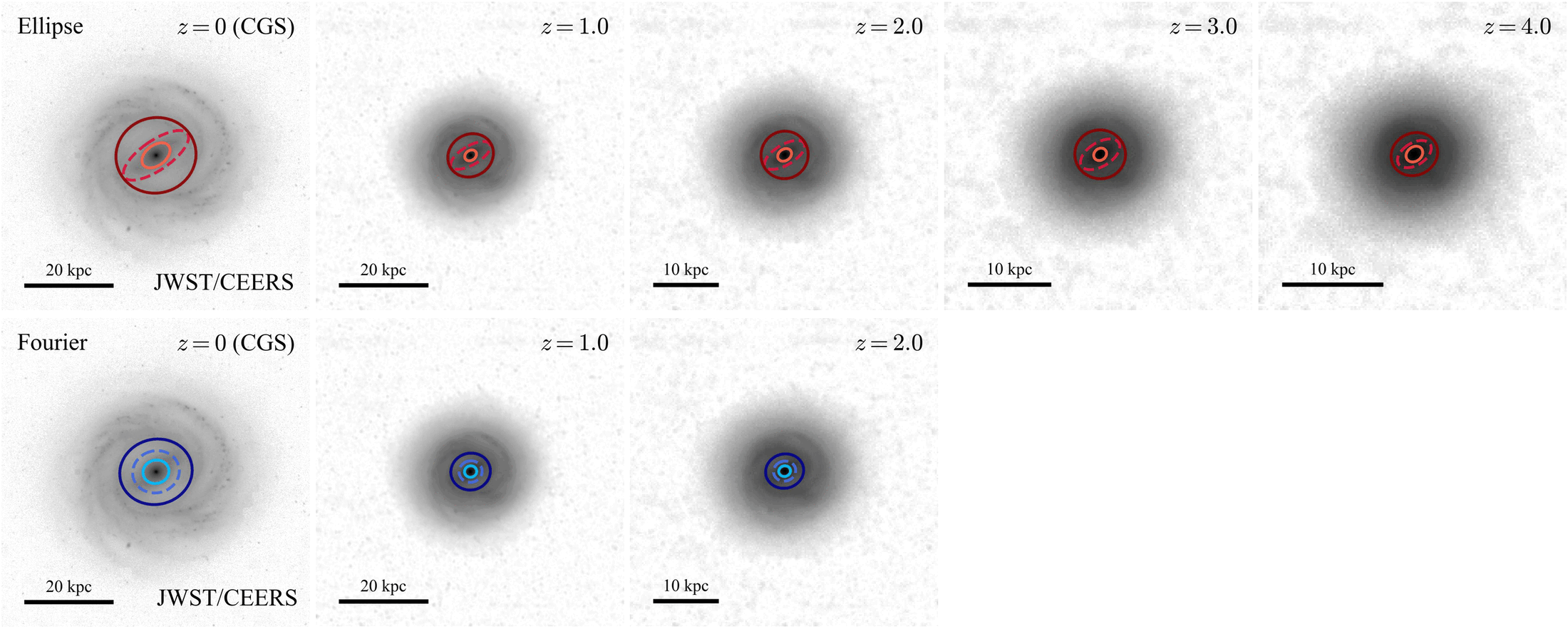}
\caption{Visual inspection of bar detection in NGC\,3450 as a function of redshift, comparing the performance of the ellipse and Fourier methods using the high-redshift bar criteria. The mock observations pertain to HST/CANDELS up to $z=3.0$ (top two rows) and to JWST/CEERS up to $z=4.0$ (bottom two rows). The first column shows the original $z=0$ CGS image for reference. The first and third rows display the results from the ellipse method; the crimson dashed ellipse marks the measured bar length ($R_{\rm bar}$), and the tomato/dark red solid ellipses mark the inner and outer boundaries. This method accurately traces the bar geometry even at high redshifts (e.g., $z=4.0$ in CEERS). The second and fourth rows give the results for the Fourier method; the royal blue dashed ellipse indicates the peak of the $I_2/I_0$ profile. In the HST/CANDELS mocks, the Fourier method can misidentify broad, outer $m=2$ structures (such as spiral arms) as the bar at high redshift, leading to an overestimate of the bar size. In the JWST/CEERS mocks, the same failure mode is less severe at $z=2.0$, but the Fourier method no longer reports a robust detection at the highest redshifts shown. Blank cells indicate cases where the pipeline did not successfully report a bar detection. The nonblank images are displayed with an asinh stretch to reduce saturation in the central regions. Each image is scaled independently for display and therefore does not provide a direct comparison of surface brightness between redshifts or surveys.}
\label{fig:ngc3450_z_evolution}
\end{figure*}

To visualize the robustness of our adaptive strategy and the specific behaviors of the two detection methods, Figure~\ref{fig:ngc3450_z_evolution} presents a redshift evolution sequence for the galaxy NGC\,3450. The visual comparison reveals a distinct difference in performance. The ellipse method (first and third rows) proves remarkably robust, accurately tracing the visually apparent bar structure (marked by crimson dashed ellipses) out to $z=2.0$ in HST/CANDELS simulations and extending to $z=4.0$ in the higher quality JWST/CEERS simulations. In contrast, the Fourier method (second and fourth rows), while successful at lower redshifts, exhibits characteristic failure modes at high redshift. In the HST/CANDELS sequence, the Fourier algorithm identifies significant $m=2$ power at $z \ge 2.0$ but assigns the bar length to a much larger structure (marked by royal blue dashed ellipses), likely corresponding to the outer spiral arms or a pseudo-ring, instead of the bar itself. In the JWST/CEERS sequence, the Fourier measurement remains closer to the central bar at $z=2.0$ but no longer yields a robust detection in the highest redshift frames. This visual evidence confirms the quantitative trend observed in Section~\ref{subsubsec:quantifying_bias}, where the Fourier method tends to overestimate bar lengths or lose reliable detections at high redshift. Consequently, although the Fourier method can technically report a ``detection'' in some degraded images, its reliability in correctly isolating the bar component diminishes significantly in the regime of low signal-to-noise ratio compared to the ellipse method. Furthermore, the superior sensitivity of JWST is evident. The ellipse method successfully recovers the bar in the CEERS simulation at $z=4.0$ (Figure~\ref{fig:ngc3450_z_evolution}, bottom two rows), whereas the corresponding detection fails in the CANDELS simulation beyond $z=2.0$ (Figure~\ref{fig:ngc3450_z_evolution}, top two rows).

\begin{deluxetable*}{cccccccc}[!htbp]
\tabletypesize{\footnotesize}
\tablecolumns{8}
\tablecaption{Minimum Detectable Intrinsic Bar Length \label{tab:min_detectable_length}}
\tablewidth{0pt}
\tablehead{
    \colhead{Survey} & \colhead{$z$} & \colhead{Minimum Length} & \colhead{Scale} & \colhead{Minimum Length} & \colhead{Filter} & \colhead{PSF FWHM} & \colhead{Ratio} \\
    \colhead{} & \colhead{} & \colhead{(kpc)} & \colhead{(kpc/$\arcsec$)} & \colhead{(\arcsec)} & \colhead{} & \colhead{(\arcsec)} & \colhead{} \\
    \colhead{(1)} & \colhead{(2)} & \colhead{(3)} & \colhead{(4)} & \colhead{(5)} & \colhead{(6)} & \colhead{(7)} & \colhead{(8)}
}
\startdata
CANDELS & 0.5 & 0.69 & 5.915 & 0.117 & F814W & 0.110 & 1.06 \\
CANDELS & 1.0 & 0.61 & 7.821 & 0.078 & F125W & 0.180 & 0.44 \\
CANDELS & 2.0 & 0.96 & 8.243 & 0.116 & F160W & 0.190 & 0.61 \\
CANDELS & 3.0 & 0.46 & 7.616 & 0.060 & F160W & 0.190 & 0.32 \\
\hline
CEERS   & 1.0 & 0.61 & 7.821 & 0.078 & F115W & 0.066 & 1.19 \\
CEERS   & 2.0 & 0.52 & 8.243 & 0.063 & F200W & 0.077 & 0.81 \\
CEERS   & 3.0 & 0.46 & 7.616 & 0.060 & F277W & 0.123 & 0.49 \\
CEERS   & 4.0 & 0.42 & 6.889 & 0.060 & F356W & 0.142 & 0.43 \\
CEERS   & 5.0 & 0.39 & 6.234 & 0.062 & F410M & 0.155 & 0.40 \\
CEERS   & 6.0 & 0.36 & 5.673 & 0.064 & F444W & 0.161 & 0.40 \\
\enddata
\tablecomments{Col. (1): Survey name. Col. (2): Mock redshift. Col. (3): Minimum detected intrinsic bar length (kpc). Col. (4): Physical scale at the given redshift. Col. (5): Minimum detected intrinsic bar length (arcsec). Col. (6): Filter used for detection. Col. (7): PSF FWHM of the filter. Col. (8): Ratio of minimum bar size to PSF FWHM. The minima are evaluated for successful ellipse-method detections.}
\end{deluxetable*}

\begin{figure*}[ht!]
\centering
\includegraphics[width=\textwidth]{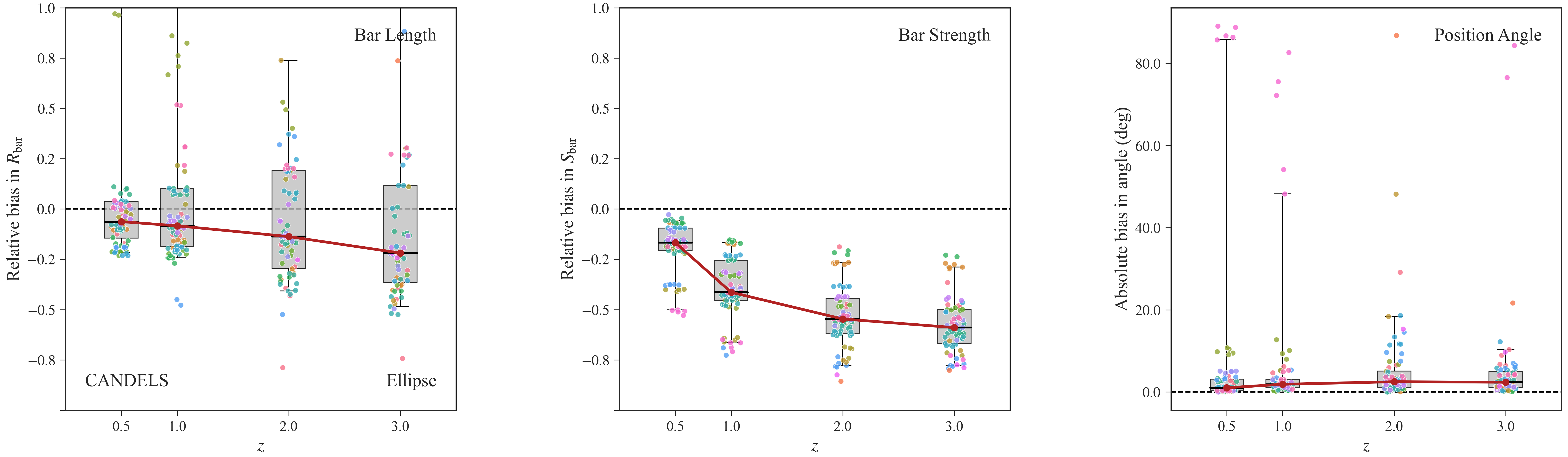} \\
\vspace{0.2cm}
\includegraphics[width=\textwidth]{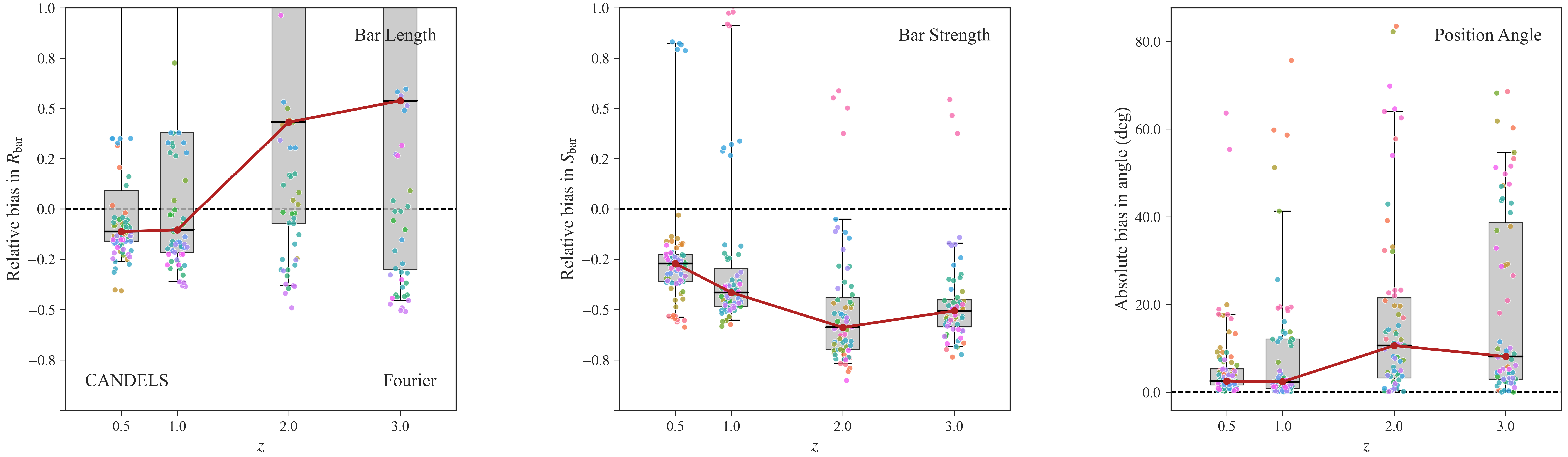}
\caption{Dependence on redshift of the measurement bias based on simulated HST/CANDELS images, for the ellipse (top) and Fourier (bottom) method. Each panel displays the bias for a different property. The $y$-axis shows the relative bias for bar length [$(R_{\rm measured} - R_{\rm true})/R_{\rm true}$] and strength, and the absolute bias for the orientation angle. The orientation angle is $\Theta_{\rm bar}$ for ellipse fitting and $\Theta_{2,\rm bar}$ for Fourier analysis. The ellipse lengths are deprojected following \citet{li_carnegie-irvine_2011}, whereas the Fourier lengths are measured from the semi-major-axis coordinates of the fixed-geometry elliptical annuli and require no additional deprojection. The black dashed line at $y=0$ indicates a perfect, unbiased measurement. The results based on the ellipse method (top) reveal a systematic negative bias (underestimation) for bar length that increases with redshift, consistent with surface brightness dimming effects. In contrast, the Fourier method (bottom) shows a positive median bias (overestimation) for bar length at $z>1.0$, indicating a tendency to overestimate bar sizes in the lower signal-to-noise regime.}
\label{fig:bias_evolution_candels}
\end{figure*}

\begin{figure*}[ht!]
\centering
\includegraphics[width=\textwidth]{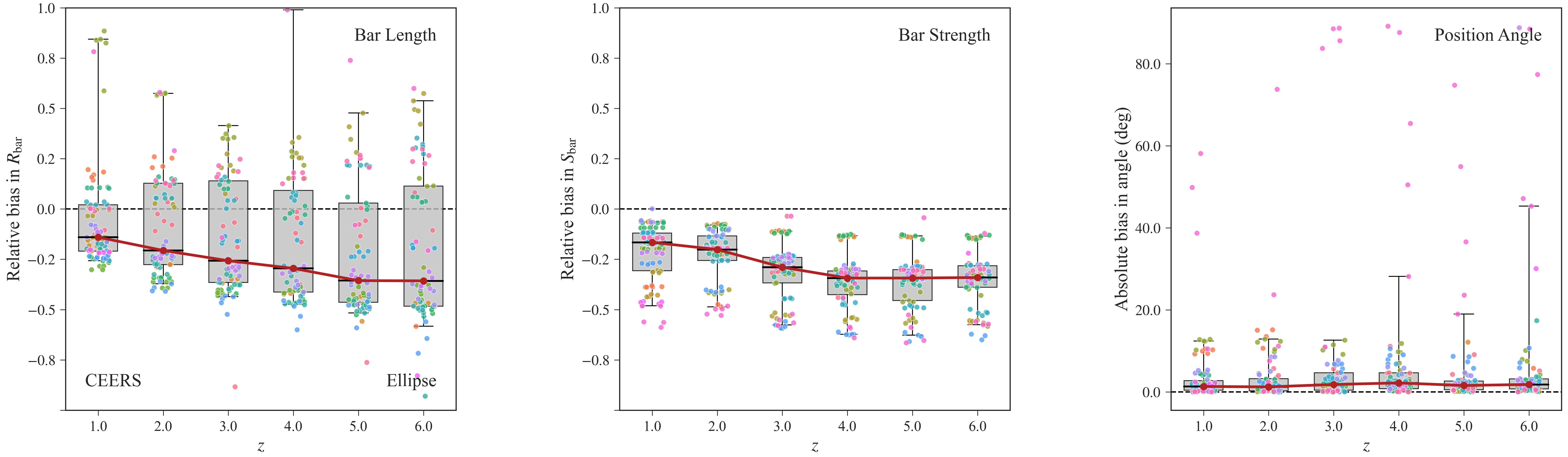} \\
\vspace{0.2cm}
\includegraphics[width=\textwidth]{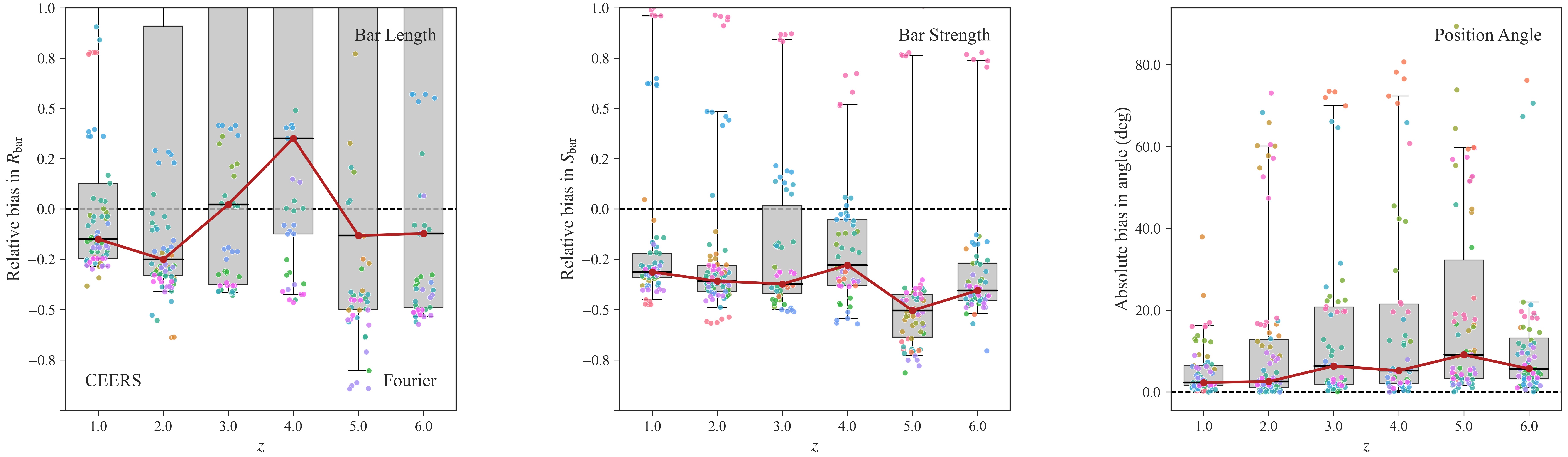}
\caption{Same as Figure~\ref{fig:bias_evolution_candels}, but for simulated JWST/CEERS images. The ellipse method (top) yields modest negative median biases in bar length out to $z \approx 3$, with larger underestimation at higher redshift. The bar strength shows a systematic underestimation. The Fourier method (bottom) shows positive median length biases at $z=3$ and 4, but the sign of the bias varies across the full redshift range.}
\label{fig:bias_evolution_ceers}
\end{figure*}

\subsection{Impact of Redshift-dependent Observational Effects on Bar Measurement}
\label{subsec:measurement_evolution}

To isolate the measurement error introduced by observational degradation
(resolution and noise) from physical cosmic evolution, we compare the
measured properties with the corresponding measurements from
the high-quality $z=0$ CGS images, using the same method. These CGS
measurements are method-dependent and have their own uncertainties.
The CGS bar length for each galaxy is evolved to the target
redshift according to the galaxy's size evolution parameter $\beta$
(Section~\ref{subsec:mock_generation}):

\begin{equation}
R_{\rm bar}^{\rm intrinsic}(z)
=
R_{\rm bar}^{\rm CGS} \times (1+z)^{\beta}.
\end{equation}

The CGS bar strengths and orientations are used directly and
are not scaled by the size-evolution factor.
Analyzing the CGS bar lengths of galaxies successfully detected
by the ellipse method allows us to identify the minimum intrinsic bar
length recovered at each redshift for both surveys
(Table~\ref{tab:min_detectable_length}). We find that the minimum
detectable intrinsic bar length corresponds to approximately
$0.3-1.2$ times the PSF FWHM at the observed wavelength.
For the higher quality JWST/CEERS simulations,
this ratio changes from 1.19 at $z=1.0$ to 0.40 at $z=6.0$.
In the HST/CANDELS simulations,
the ratio similarly drops to 0.32 at $z=3.0$.
This result suggests that with optimized, redshift-dependent criteria,
the effective detection limit can be pushed to scales comparable to or
even smaller than the PSF size, although completeness at these sub-PSF
scales warrants careful statistical interpretation.

\begin{figure*}[ht!]
\centering
\includegraphics[width=\textwidth]{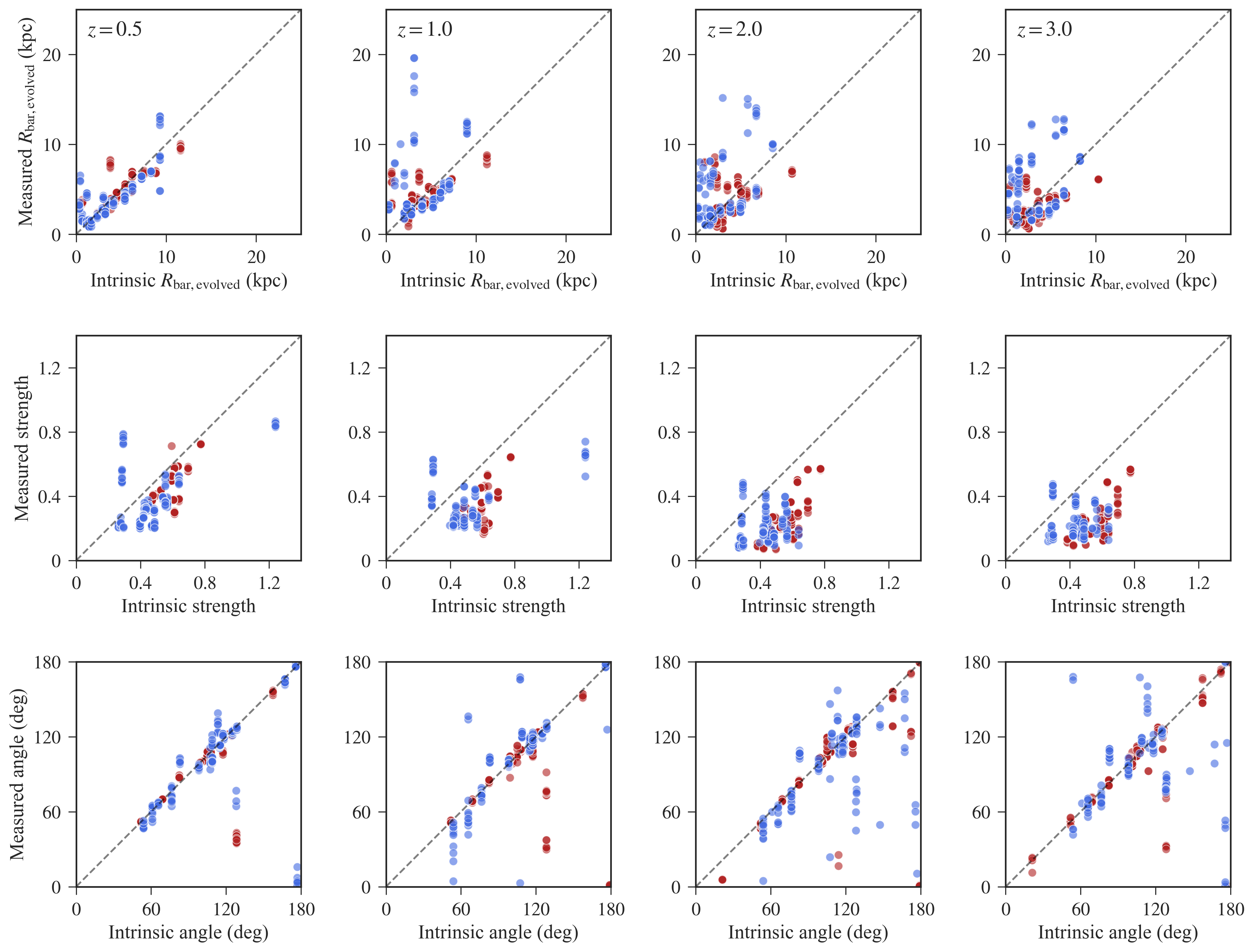}
\caption{Comparison of measured and intrinsic bar properties for HST/CANDELS simulations as a function of redshift, for bar length (top), bar strength (middle), and orientation angle ($\Theta_{\rm bar}$ for ellipse fitting and $\Theta_{2,\rm bar}$ for Fourier analysis; bottom). Red points represent the ellipse method, and blue points represent the Fourier method. The ellipse method has smaller scatter in bar length and strength (Table~\ref{tab:bias_statistics}). The dashed line shows the 1:1 relation.}
\label{fig:scatter_all_candels}
\end{figure*}

\begin{figure*}[ht!]
\centering
\includegraphics[width=\textwidth]{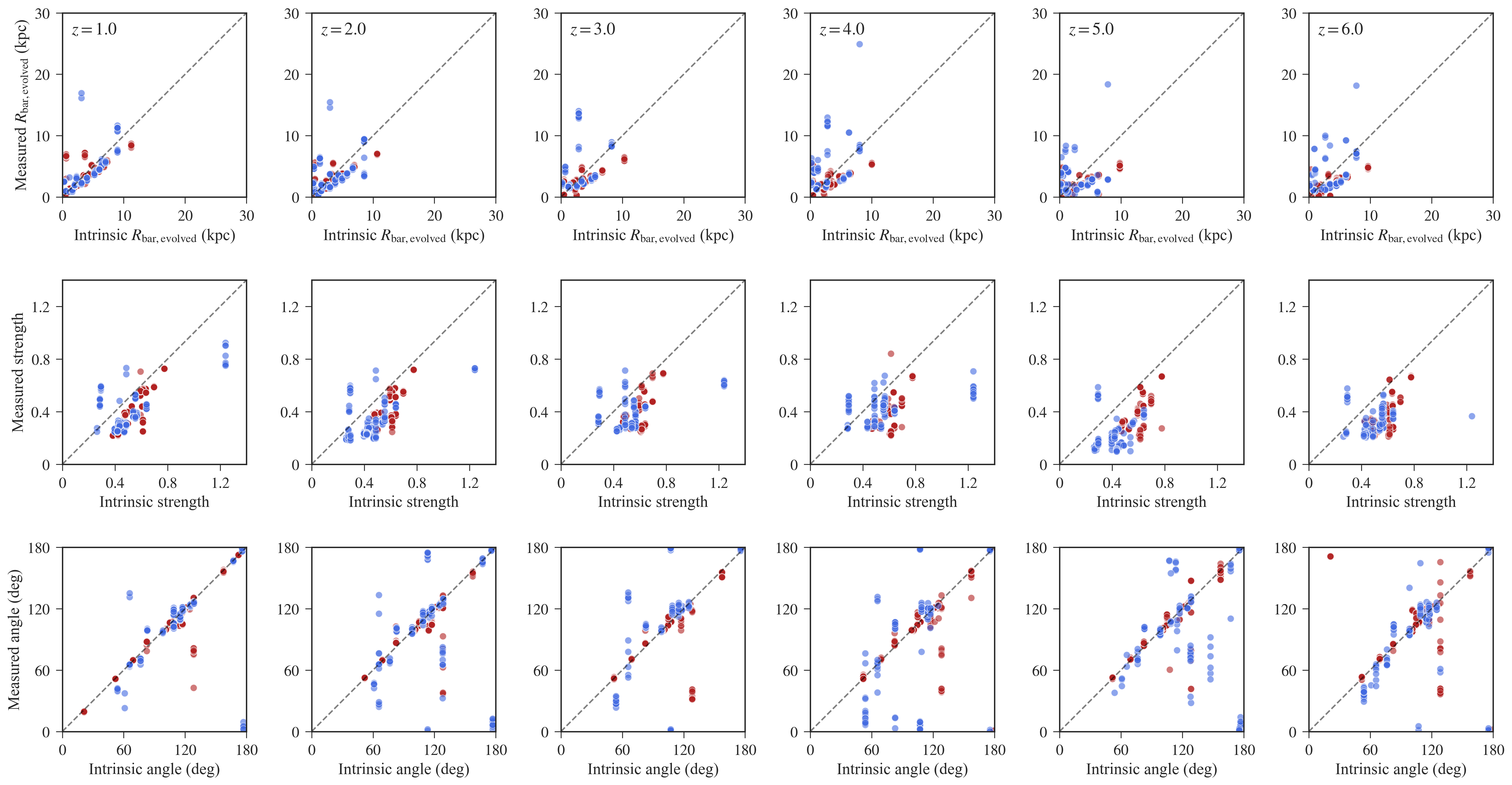}
\caption{Same as Figure~\ref{fig:scatter_all_candels}, but for simulated JWST/CEERS images. The ellipse method has smaller scatter in bar length and strength at all simulated redshifts through $z=6.0$.}
\label{fig:scatter_all_ceers}
\end{figure*}

\begin{figure*}[p]
\centering
\includegraphics[width=\textwidth]{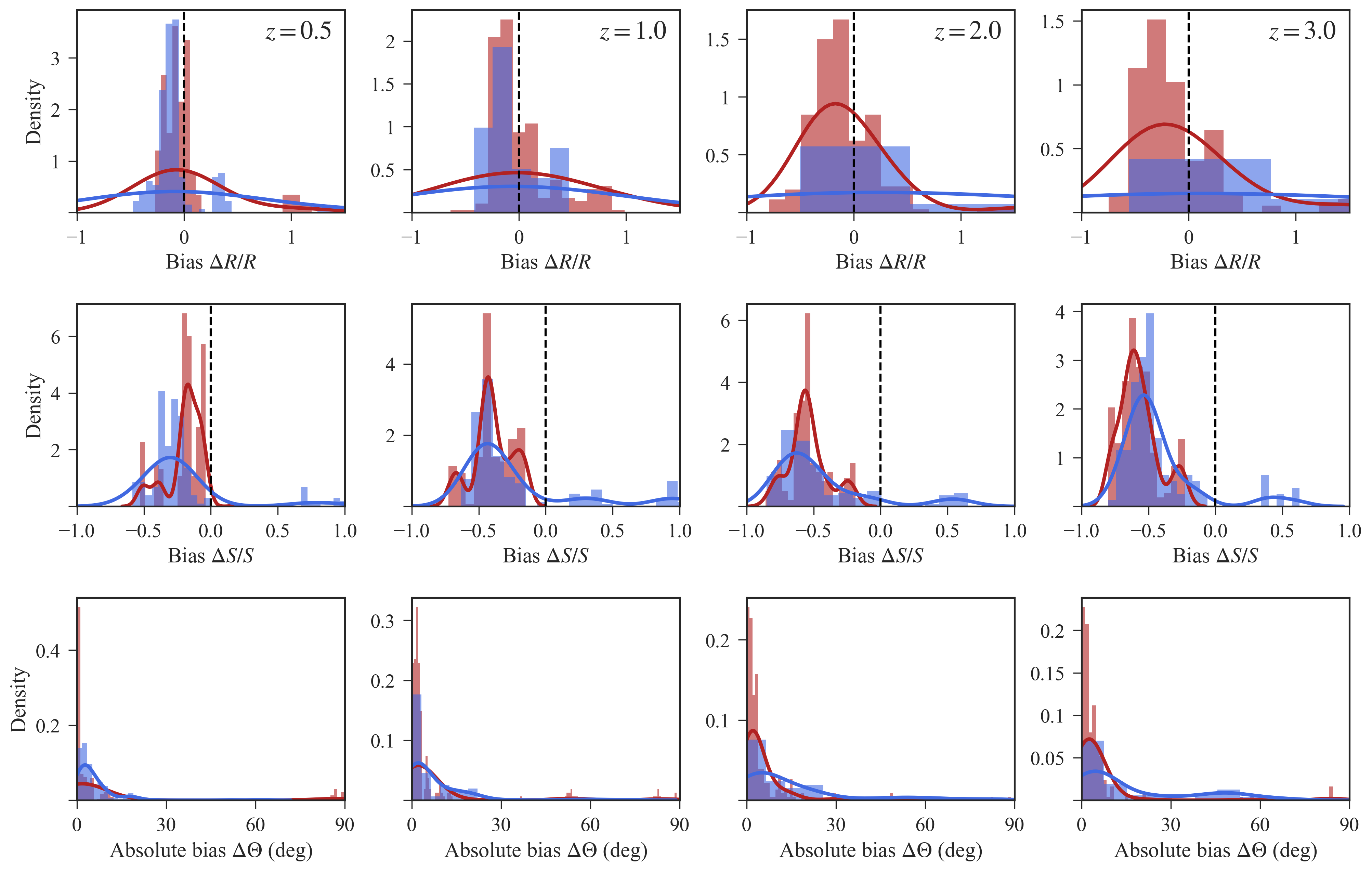}
\caption{Distribution of measurement bias for HST/CANDELS simulations, for relative bias in bar length (top), relative bias in bar strength (middle), and absolute bias in orientation angle (bottom). The ellipse method (red) has smaller standard deviations in bar length and strength than the Fourier method (blue), while the comparison of orientation scatter depends on redshift (Table~\ref{tab:bias_statistics}).}
\label{fig:hist_all_candels}
\end{figure*}

\begin{figure*}[p]
\centering
\includegraphics[width=\textwidth]{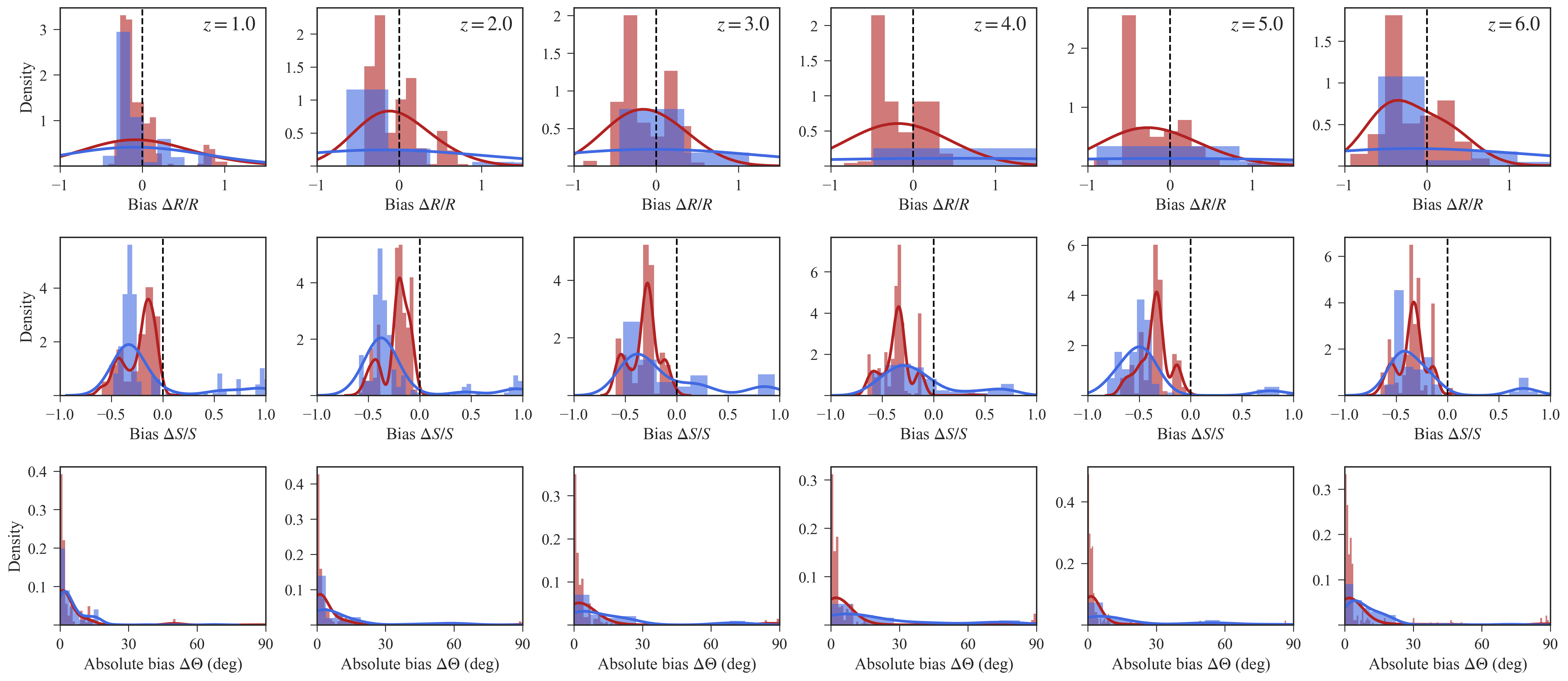}
\caption{Same as Figure~\ref{fig:hist_all_candels}, but for JWST/CEERS simulations. The ellipse method (red) has smaller scatter in bar length and strength than the Fourier method (blue). The relative orientation scatter depends on redshift.}
\label{fig:hist_all_ceers}
\end{figure*}

To assess measurement accuracy, we analyze the deviation of measured
values from intrinsic values for bar length ($R_{\rm bar}$), strength
($S_{\rm bar}$), and orientation angle
($\Theta_{\rm bar}$ for ellipse fitting and
$\Theta_{2,\rm bar}$ for Fourier analysis). Note that bar strength is
defined as the maximum ellipticity ($e_{\rm max}$) for the ellipse method
and the maximum normalized $m=2$ Fourier amplitude
[$(I_2/I_0)_{\rm max}$] for the Fourier method. Figures
\ref{fig:bias_evolution_candels} and \ref{fig:bias_evolution_ceers}
visualize the overall evolution of measurement bias across the redshift
range. A clear difference exists between the two methods regarding bar
length. The ellipse method exhibits a systematic negative median bias,
underestimating bar lengths by roughly $4\%-36\%$, depending
on redshift (Table~\ref{tab:bias_statistics}). This is physically
expected, as surface brightness dimming causes the faint outer isophotes
of the bar to become indistinguishable from the disk background.
The sample standard deviation of the fractional length bias
ranges from 0.88 to 2.22. By comparison, the Fourier measurements show
substantially larger scatter, with standard deviations from 2.40 to
6.14, while their median length bias can have either sign.
Figures \ref{fig:scatter_all_candels} and \ref{fig:scatter_all_ceers}
compare all parameters, Figures \ref{fig:hist_all_candels} and
\ref{fig:hist_all_ceers} show the detailed distribution of biases, and
Table~\ref{tab:bias_statistics} gives the median bias and the
sample standard deviation of each distribution.

In terms of bar length, the ellipse method (red points and histograms)
demonstrates superior reliability driven by its relatively higher precision
compared to Fourier analysis. While bar length measurements exhibit a
systematic negative bias that tends to persist at higher redshifts
(e.g., reaching $-36\%$ at $z=5.0$ in CEERS), this
underestimation is a consistent consequence of surface brightness
dimming eroding the faint outer isophotes of the bar. In contrast, the
Fourier method shows substantially broader distributions. A striking
example is found in the CEERS simulation at $z=4.0$, where the Fourier
method shows a positive median bias ($+0.55$) with a standard
deviation of 6.14. Contamination from larger-scale spiral
arms or outer rings, as illustrated in
Figure~\ref{fig:ngc3450_z_evolution}, can contribute to such large
positive errors.

For bar strength, both methods show
negative median biases in every redshift bin, consistent with
PSF blurring, which dilutes the peak density of the bar relative to the
surrounding disk. The ellipse method also yields smaller sample
standard deviations in bar strength.

For orientation, the ellipse method yields median absolute differences
from $1.0^\circ$ to $2.3^\circ$, compared with
from $2.6^\circ$ to $12.5^\circ$ for the Fourier method
(Table~\ref{tab:bias_statistics}). The corresponding sample standard
deviations range from $10.6^\circ$ to $23.3^\circ$ for
ellipse fitting and from $8.0^\circ$ to $25.6^\circ$ for
Fourier analysis. Thus, although the median absolute
orientation differences are smaller for ellipse fitting, neither method
has smaller orientation scatter in every redshift bin.

\begin{deluxetable*}{cclrcr}[!ht]
\tabletypesize{\footnotesize}
\tablecolumns{6}
\tablecaption{Measurement Bias Statistics \label{tab:bias_statistics}}
\tablewidth{0pt}
\tablehead{
    \colhead{Survey} & \colhead{$z$} & \colhead{Method} & \colhead{Bias $R_{\rm bar}$} & \colhead{Bias $S_{\rm bar}$} & \colhead{Bias Angle (deg)} \\
    \colhead{(1)} & \colhead{(2)} & \colhead{(3)} & \colhead{(4)} & \colhead{(5)} & \colhead{(6)}
}
\startdata
CANDELS & 0.5 & Ellipse & $-0.04 \pm 1.16$ & $-0.17 \pm 0.13$ & $1.0 \pm 23.3$ \\
        &     & Fourier & $-0.10 \pm 2.44$ & $-0.27 \pm 0.50$ & $3.4 \pm 8.0$ \\
CANDELS & 1.0 & Ellipse & $-0.07 \pm 2.22$ & $-0.42 \pm 0.15$ & $2.0 \pm 17.7$ \\
        &     & Fourier & $-0.09 \pm 3.03$ & $-0.40 \pm 0.45$ & $2.6 \pm 12.7$ \\
CANDELS & 2.0 & Ellipse & $-0.15 \pm 0.88$ & $-0.55 \pm 0.15$ & $2.3 \pm 10.6$ \\
        &     & Fourier & $0.27 \pm 4.26$ & $-0.57 \pm 0.37$ & $7.1 \pm 19.3$ \\
CANDELS & 3.0 & Ellipse & $-0.21 \pm 1.28$ & $-0.59 \pm 0.15$ & $2.3 \pm 13.7$ \\
        &     & Fourier & $0.04 \pm 4.78$ & $-0.51 \pm 0.29$ & $6.8 \pm 19.7$ \\
\hline
CEERS   & 1.0 & Ellipse & $-0.13 \pm 1.91$ & $-0.17 \pm 0.15$ & $1.3 \pm 10.7$ \\
        &     & Fourier & $-0.16 \pm 2.40$ & $-0.31 \pm 0.45$ & $2.9 \pm 8.3$ \\
CEERS   & 2.0 & Ellipse & $-0.15 \pm 1.14$ & $-0.20 \pm 0.13$ & $1.2 \pm 11.1$ \\
        &     & Fourier & $-0.25 \pm 3.58$ & $-0.37 \pm 0.41$ & $3.1 \pm 18.3$ \\
CEERS   & 3.0 & Ellipse & $-0.17 \pm 1.21$ & $-0.30 \pm 0.14$ & $1.8 \pm 20.3$ \\
        &     & Fourier & $0.09 \pm 3.57$ & $-0.32 \pm 0.44$ & $8.3 \pm 20.1$ \\
CEERS   & 4.0 & Ellipse & $-0.27 \pm 1.69$ & $-0.35 \pm 0.14$ & $2.1 \pm 18.6$ \\
        &     & Fourier & $0.55 \pm 6.14$ & $-0.23 \pm 0.36$ & $12.5 \pm 25.6$ \\
CEERS   & 5.0 & Ellipse & $-0.36 \pm 1.50$ & $-0.34 \pm 0.14$ & $1.4 \pm 11.1$ \\
        &     & Fourier & $-0.19 \pm 5.40$ & $-0.48 \pm 0.37$ & $7.8 \pm 24.1$ \\
CEERS   & 6.0 & Ellipse & $-0.27 \pm 0.92$ & $-0.34 \pm 0.14$ & $1.8 \pm 17.3$ \\
        &     & Fourier & $-0.23 \pm 4.01$ & $-0.40 \pm 0.37$ & $6.0 \pm 12.1$ \\
\enddata
\tablecomments{Col. (1): Survey name. Col. (2): Mock redshift. Col. (3): Measurement method. Col. (4): Relative bias in bar length, $(R_{\rm measured} - R_{\rm true})/R_{\rm true}$. Col. (5): Relative bias in bar strength. Col. (6): Absolute orientation difference, using $\Theta_{\rm bar}$ for ellipse fitting and $\Theta_{2,\rm bar}$ for Fourier analysis. Each entry gives the median followed by the sample standard deviation of the distribution.}
\end{deluxetable*}

\section{Discussion} \label{sec:discussion}

\subsection{Impact of Observational Effects and Bias Correction}
\label{subsec:bias_correction}

Our analysis unequivocally demonstrates that redshift-dependent observational effects---specifically surface brightness dimming and resolution degradation---introduce significant biases in both the detection and measurement of galactic bars. These effects, if left uncorrected, can lead to erroneous conclusions about the cosmic evolution of bar fraction and bar properties. Based on our quantitative results (Section~\ref{sec:results}), we summarize the biases and recommended correction strategies below.

{\it{Bar detection and fraction ($f_{\rm bar}$):}}
Using static, local detection criteria at high redshift results in a severe underestimation of the bar fraction. For HST/CANDELS, this ``missing bar'' problem becomes critical at $z > 0.5$, where the local ellipse criteria miss over 50\% of input bars. We strongly recommend adopting redshift-dependent adaptive criteria, as detailed in Table~\ref{tab:optimized_criteria}. By relaxing ellipticity thresholds to account for PSF blurring, the ellipse recovery rate remains above $60\%$ out to $z=3$ with HST and $z=6$ with JWST. The JWST Fourier recovery rates range from $47\%$ to $76\%$.

Regarding the physical resolution limit, our optimized criteria allow for the recovery of bars significantly smaller than previously assumed. As shown in Table~\ref{tab:min_detectable_length}, the minimum detectable intrinsic bar length corresponds to approximately $0.3-1.2$ times the PSF FWHM at the observed wavelength. This suggests that with rigorous, adaptive criteria, the effective detection limit can be pushed to scales comparable to or slightly smaller than the PSF size, although completeness at these scales warrants careful individual verification.

{\it{Bar length ($R_{\rm bar}$):}}
The ellipse method systematically underestimates bar length at high redshift because surface brightness dimming renders the faint outer isophotes of the bar indistinguishable from the disk background. As demonstrated in Table~\ref{tab:bias_statistics}, this negative bias ranges from $\sim 4\%$ to $\sim 36\%$. While the scatter in the measurements ($\sigma = 0.88$--$2.22$) indicates non-negligible variation, the ellipse method remains more stable than the Fourier method, which shows substantially larger scatter ($\sigma = 2.40$--$6.14$) and can overestimate bar lengths when contaminated by spiral arms or other outer structures. To facilitate future studies, we provide recommended correction factors ($f_{R}$) for the ellipse method in Table~\ref{tab:correction_factors}, with which the intrinsic bar length can be estimated as $R_{\rm intrinsic} \approx R_{\rm measured} \times f_{R}$.

{\it{Bar strength and orientation:}}
Bar strength is underestimated in every redshift bin, as PSF convolution dilutes the peak density contrast of the bar relative to the surrounding disk. Similar to bar length, the ellipse method maintains a better correlation with intrinsic values compared to the Fourier method. We provide corresponding correction factors ($f_{S}$) in Table~\ref{tab:correction_factors} and Equation~\ref{eq:correction_funcs}. Regarding orientation, the ellipse method yields robust measurements with median absolute differences below $3^\circ$ in all sampled redshift bins; therefore, we do not derive an orientation correction.

\begin{deluxetable*}{cccccc}[!htbp]
\tabletypesize{\footnotesize}
\tablecolumns{6}
\tablecaption{Recommended Correction Factors for Bar Properties Measured Using the Ellipse Method \label{tab:correction_factors}}
\tablewidth{0pt}
\tablehead{
    \colhead{Survey} & \colhead{$z$} & \multicolumn{2}{c}{Bar Length ($R_{\rm bar}$)} & \multicolumn{2}{c}{Bar Strength ($S_{\rm bar}$)} \\
    \colhead{} & \colhead{} & \colhead{Bias ($\Delta R/R$)} & \colhead{Factor ($f_{R}$)} & \colhead{Bias ($\Delta S/S$)} & \colhead{Factor ($f_{S}$)} \\
    \colhead{(1)} & \colhead{(2)} & \colhead{(3)} & \colhead{(4)} & \colhead{(5)} & \colhead{(6)}
}
\startdata
HST/CANDELS & 0.5 & $-0.04$ & $1.05$ & $-0.17$ & $1.20$ \\
            & 1.0 & $-0.07$ & $1.08$ & $-0.42$ & $1.71$ \\
            & 2.0 & $-0.15$ & $1.18$ & $-0.55$ & $2.24$ \\
            & 3.0 & $-0.21$ & $1.26$ & $-0.59$ & $2.46$ \\
\hline
JWST/CEERS  & 1.0 & $-0.13$ & $1.15$ & $-0.17$ & $1.20$ \\
            & 2.0 & $-0.15$ & $1.18$ & $-0.20$ & $1.25$ \\
            & 3.0 & $-0.17$ & $1.20$ & $-0.30$ & $1.42$ \\
            & 4.0 & $-0.27$ & $1.38$ & $-0.35$ & $1.54$ \\
            & 5.0 & $-0.36$ & $1.57$ & $-0.34$ & $1.52$ \\
            & 6.0 & $-0.27$ & $1.37$ & $-0.34$ & $1.52$ \\
\enddata
\tablecomments{Col. (1): Survey name. Col. (2): Mock redshift. Col. (3): Median relative bias in bar length from Table~\ref{tab:bias_statistics}. Col. (4): Correction factor for bar length, $f_{R} = 1 / (1 + \text{Bias}_{R})$. Col. (5): Median relative bias in bar strength from Table~\ref{tab:bias_statistics}. Col. (6): Correction factor for bar strength, $f_{S} = 1 / (1 + \text{Bias}_{S})$.}
\end{deluxetable*}

To enable continuous corrections across the probed redshift range, we provide the following empirical relations for the ellipse-method correction factors:

\begin{equation}
\label{eq:correction_funcs}
\begin{aligned}
    f_{R, \rm CANDELS}(z) &= 0.09z + 1.00 \\
    f_{S, \rm CANDELS}(z) &= 0.49z + 1.11 \\
    f_{R, \rm CEERS}(z) &= 0.07z + 1.06 \\
    f_{S, \rm CEERS}(z) &= 0.07z + 1.16.
\end{aligned}
\end{equation}

\noindent Given the limited number of discrete redshift snapshots in our mock catalog, we adopt a conservative linear least-squares fit to characterize the evolution of the bias. This first-order approximation avoids overfitting to the sparse data points while capturing the overall trend of increasing bias with redshift. Note that for CEERS, the bias in bar strength grows significantly slower with redshift (slope $\sim 0.07$) compared to CANDELS (slope $\sim 0.49$), reflecting the superior resolution of JWST at high redshifts. We recommend applying these functions within the redshift ranges simulated in this work ($z=0.5-3.0$ for CANDELS and $z=1.0-6.0$ for CEERS).

\subsection{Practical Recommendations for Future Surveys}
\label{subsec:recommendations}

Based on the performance comparison of the two methods, we propose a standardized workflow for analyzing bars in large high-redshift galaxy surveys, such as those conducted with HST and JWST, as well as upcoming missions such as the Chinese Space Station Survey Telescope (CSST; \citealt{Gong_2026}). We recommend the ellipse method as the primary tool for both bar detection and measurement of its basic photometric properties. While the Fourier method is highly sensitive to non-axisymmetric signals, it can confuse spiral arms or outer rings with the bar component in the regime of low signal-to-noise ratio (as visualized in Figure~\ref{fig:ngc3450_z_evolution}). The ellipse method, constrained by the geometric alignment of isophotes, proves to be far more robust against such contamination.

Moreover, accurate interpretation requires a reference set of mock simulations. One should not rely on generic corrections from the literature. Instead, we recommend generating mock observations by artificially redshifting a local sample of barred galaxies together with an appropriate negative-control sample matched to the specific PSF and depth of the target survey, while reproducing the luminosity and size distribution of the galaxy sample under consideration. Such mocks are critical for optimizing detection criteria---maximizing the true positive rate while minimizing false positives relative to the adopted control classification---and for deriving survey-specific bias correction factors. Once the criteria are optimized and detections are made, the results from the mock reference set serve two additional critical purposes: deriving bias correction factors for the raw measurements and estimating realistic error bars using the standard deviation of the mock bias distribution.

\subsection{The 2D Multi-component Decomposition Approach}
\label{subsec:galfit_limitations}

An alternative approach for bar analysis is to employ 2D multi-component decomposition (e.g., using \texttt{GALFIT}; \citealt{peng_galfit_2002,peng_galfit_2010}) to model the galaxy as a superposition of distinct physical components. We chose not to adopt this method for our automated pipeline due to the severe degeneracies inherent in parametric fitting of high-redshift galaxy images.

As shown by \citet{erwin_composite_2021} for the nearby barred
galaxies NGC\,4608 and NGC\,4643, simple bulge--disk decompositions can
lead to an overestimate of the bulge contribution when light from bars
and other central structures is not modeled separately. These results
highlight the importance of adequate component modeling in barred
galaxies.

In contrast, our isophotal analysis methods are non-parametric. They measure the signal directly from the image structure without assuming a specific analytical profile, making them more robust for determining the presence and extent of bars in high-redshift galaxies, which are often irregular or clumpy (e.g., \citealt{Guo_2012, Kalita_2024}). While 2D decomposition remains a powerful tool for detailed structural analysis, applying it blindly to large high-redshift samples carries significant risks. A dedicated quantification of how the unrecovered bar flux biases the bulge and disk scaling relations in high-$z$ surveys will be the subject of our forthcoming work.

\subsection{Caveats and Limitations of the Mock-based Approach}
\label{subsec:caveats}

While our mock-based analysis provides a rigorous calibration for resolution and surface brightness dimming effects, the applicability of our derived corrections and detection limits is subject to caveats inherent to the base sample and the assumed galaxy morphologies.

The representativeness of the base catalog is a key source of uncertainty. Our simulations rely on the CGS sample, which is a magnitude-limited survey of bright ($B_T < 12.9$ mag), local galaxies. Consequently, the parameter space at the low-mass end ($M_{\star} < 10^{10}\,M_\odot$) is sampled poorly in our base catalog. This implies that our mock predictions for low-mass galaxies rely on a limited set of morphological templates, which may not fully capture the structural variance of the dwarf galaxy population. Furthermore, although we explicitly incorporate redshift-dependent luminosity and size evolution (Section~\ref{sec:style}), this procedure assumes that high-redshift galaxies are structurally homologous to local spirals. In reality, high-redshift disk galaxies often exhibit intrinsically different morphologies, characterized by higher gas fractions, clumpier star formation, and higher turbulence compared to the settled thin disks in the local Universe \citep[e.g.,][]{sheth_evolution_2008, Guo_2012}. Therefore, the bar recovery rates derived here should be interpreted as the recoverability of evolved, settled bars placed at high redshift. If primordial bars in gas-rich, turbulent disks have fundamentally different surface brightness profiles, our corrections may need further refinement using hydrodynamical simulations. The $UVJ$ star-forming selection used to calibrate luminosity and size evolution is likewise only a statistical proxy for disk morphology; contamination by non-disk systems and omission of quiescent disks are additional sources of uncertainty.

The impact of galaxy interactions also warrants consideration. Our false-positive analysis relies on a control sample of galaxies classified as unbarred in CGS. As discussed in Section~\ref{subsec:sample_selection}, comparison with available mid-infrared and NIR classifications indicates weak, intermediate, or nuclear bar-like structures in several objects, and thus our false-positive rates should be interpreted relative to the adopted optical control classification. In addition, the galaxy merger rate increases significantly with redshift (\citealt{conselice_structures_2009}). Interacting systems often display strong tidal tails, warps, or distorted spiral arms that could mimic the photometric signature of a bar. Dedicated mock simulations should be performed to quantify the degree to which galaxies with such distorted morphologies can be recognized and therefore excluded from any specific bar study.

\section{Summary}
\label{sec:summary}

We present a comprehensive analysis of how redshift-dependent observational effects bias the study of galactic bars. Utilizing a local benchmark based on high-quality, high-resolution images of nearby galaxies and generating rigorous mock observations for HST and JWST, we quantify the detection limits and measurement biases for automated methods. Our main conclusions are:

\begin{enumerate}

\item Standard criteria for detecting bars in local galaxies fail at $z > 0.5$ for HST. Adaptive, redshift-dependent criteria are essential to detect bars robustly
at high redshift. With optimized criteria, we find that the minimum detectable intrinsic bar size can be pushed to $\sim 0.3-1.2$ times the PSF FWHM, significantly extending the accessible range of bar sizes compared to static criteria.

\item JWST/CEERS significantly outperforms HST/CANDELS, pushing the reliable detection limit to smaller physical scales and higher redshifts ($z \approx 4-6$).

\item We recommend the ellipse method for automated analysis, as it outperforms the Fourier method in both reliability and accuracy at high redshift. The Fourier method shows substantially larger scatter in bar length, and contamination from spiral arms or other outer structures can produce large overestimation errors.

\item Surface brightness dimming causes bar lengths measured by the ellipse method to be underestimated systematically by $\sim 4\%-36\%$ depending on redshift. Bar strength is also systematically underestimated, with biases ranging from $\sim 17\%$ to $\sim 59\%$.

\item Reliable measurement and interpretation of high-redshift bar statistics require tailored mock simulations. We recommend generating mock observations by artificially redshifting local benchmark barred galaxies and an appropriate unbarred control sample to match the specific depth and PSF of a given survey. This approach is critical not only for deriving measurement correction factors but, just as importantly, for optimizing detection criteria. Criteria derived from mocks significantly outperform static local criteria, achieving the optimal balance between high completeness and low contamination relative to the adopted control sample.

\end{enumerate}

The correction strategies and practical recommendations outlined above enable future studies to mitigate effectively observational biases, paving the way for a more accurate understanding of the formation and evolution of galactic bars across cosmic time.

\begin{acknowledgments}
LCH was supported by the National Science Foundation of China (12233001) and the China Manned Space Program (CMS-CSST-2025-A09). ZYL was supported by the National Natural Science Foundation of China (12233001, 12533004), the National Key R\&D Program of China (2024YFA1611602), and the Shanghai Natural Science Research Grant (24ZR1491200). We benefitted from helpful advice from Chang-Hao Chen, Boris Kalita, Ruancun Li, Jinyi Shangguan, Wen Sun, and Si-Yue Yu.
\end{acknowledgments}

\facilities{Las Campanas Observatory du Pont Telescope,
HST (ACS, WFC3), JWST (NIRCam)}

\software{Photutils \citep{bradley_astropyphotutils_2022},
Astropy \citep{astropy_collaboration_2022},
Reproject \citep{robitaille_reproject_2020},
SciPy \citep{virtanen_scipy_2020},
Galfit \citep{peng_galfit_2002, peng_galfit_2010}
}

\appendix

\section{Literature Check of the Unbarred Control Sample}
\label{app:control-audit}
\restartappendixnumbering
\renewcommand{\theHtable}{\thesection.\arabic{table}}

The 15 control galaxies were selected using the unbarred flag in the CGS catalog of \citet{li_carnegie-irvine_2011}; our image analysis uses their CGS $R$-band images. To assess the wavelength dependence of these classifications, we cross-matched the sample against the mid-infrared morphologies published for nine galaxies in S$^4$G \citep{buta_classical_2015} and for NGC\,1317 in CS$^4$G \citep{sanchez_alarcon_cs4g_2025}. For the five remaining objects, we checked their CGS catalog properties against the S$^4$G parent-sample limits \citep{sheth_s4g_2010} and searched for independent NIR morphology information. Table~\ref{tab:control-audit} summarizes the object-by-object results.

\begin{table}[ht!]
\caption{Literature Check of the Unbarred Control Sample\label{tab:control-audit}}
\centering
{\scriptsize
\setlength{\tabcolsep}{3pt}
\renewcommand{\arraystretch}{1.10}
\begin{tabular}{c l l}
\hline\hline
Galaxy & \parbox[t]{0.54\textwidth}{Published comparison} & \parbox[t]{0.25\textwidth}{Interpretation adopted here} \\
\hline
IC\,1993 & \parbox[t]{0.54\textwidth}{S$^4$G: $(\rm R')$SA(s)ab; mean family index $\langle F\rangle=0.12$.} & \parbox[t]{0.25\textwidth}{No strong bar; weak or ambiguous family signal.} \\
IC\,2056 & \parbox[t]{0.54\textwidth}{S$^4$G: (L)SA(rs)bc; $\langle F\rangle=0.00$.} & \parbox[t]{0.25\textwidth}{No clear large-scale bar.} \\
IC\,5325 & \parbox[t]{0.54\textwidth}{S$^4$G: SA(s)bc; $\langle F\rangle=0.00$.} & \parbox[t]{0.25\textwidth}{No clear large-scale bar.} \\
IC\,5332 & \parbox[t]{0.54\textwidth}{S$^4$G: SAB(s)cd; $\langle F\rangle=0.25$.} & \parbox[t]{0.25\textwidth}{Weak or intermediate bar.} \\
NGC\,245 & \parbox[t]{0.54\textwidth}{Outside the S$^4$G/CS$^4$G velocity and distance limits ($v=4074$ km s$^{-1}$; $D=51.3$ Mpc); no homogeneous published NIR morphology classification located.} & \parbox[t]{0.25\textwidth}{CGS optical classification only; absence of a NIR classification is not evidence of bar absence.} \\
NGC\,1309 & \parbox[t]{0.54\textwidth}{S$^4$G: SAB(s)bc; $\langle F\rangle=0.25$.} & \parbox[t]{0.25\textwidth}{Weak or intermediate bar.} \\
NGC\,1317 & \parbox[t]{0.54\textwidth}{CS$^4$G: intermediate-family and nuclear-bar notation.} & \parbox[t]{0.25\textwidth}{Weak, intermediate, or nuclear bar.} \\
NGC\,5324 & \parbox[t]{0.54\textwidth}{Outside the S$^4$G/CS$^4$G velocity and distance limits ($v=3044$ km s$^{-1}$; $D=44.0$ Mpc); no homogeneous published NIR morphology classification located.} & \parbox[t]{0.25\textwidth}{CGS optical classification only.} \\
NGC\,5468 & \parbox[t]{0.54\textwidth}{S$^4$G: SAB(s)c/SAB(s)cd; $\langle F\rangle=0.25$; the table note indicates a small central oval.} & \parbox[t]{0.25\textwidth}{Weak or intermediate bar or oval.} \\
NGC\,6215 & \parbox[t]{0.54\textwidth}{Outside the S$^4$G/CS$^4$G latitude limit ($b=-9.27\degr$). The Ohio State University (OSU) $H$-band study gives SABbc; its description notes an elliptical high-surface brightness bulge but reports no other evidence of a bar \citep{eskridge_near-ir_2002}.} & \parbox[t]{0.25\textwidth}{Ambiguous weak NIR evidence, not a secure all-wavelength unbarred case.} \\
NGC\,6699 & \parbox[t]{0.54\textwidth}{Outside the S$^4$G/CS$^4$G velocity, distance, and latitude limits ($v=3347$ km s$^{-1}$; $D=45.8$ Mpc; $b=-22.65\degr$); no homogeneous published NIR classification located. CGS metadata compile SABb and SAB(rs)bc: optical types.} & \parbox[t]{0.25\textwidth}{Optical classifications conflict with the adopted \citet{li_carnegie-irvine_2011} flag.} \\
NGC\,6935 & \parbox[t]{0.54\textwidth}{Outside the S$^4$G/CS$^4$G velocity and distance limits ($v=4486$ km s$^{-1}$; $D=60.3$ Mpc); no homogeneous published NIR classification located. CGS metadata compile mixed SABa and $(\rm R')$SA(r)a optical types.} & \parbox[t]{0.25\textwidth}{Mixed optical evidence; retained only as a \citet{li_carnegie-irvine_2011} optical control.} \\
NGC\,7213 & \parbox[t]{0.54\textwidth}{S$^4$G: SA/lens-like classification; $\langle F\rangle=0.00$.} & \parbox[t]{0.25\textwidth}{No clear large-scale bar.} \\
NGC\,7371 & \parbox[t]{0.54\textwidth}{S$^4$G: SAB(rs)ab; $\langle F\rangle=0.38$.} & \parbox[t]{0.25\textwidth}{Weak or intermediate bar.} \\
NGC\,7727 & \parbox[t]{0.54\textwidth}{S$^4$G: SA; $\langle F\rangle=0.00$, with a possible small nuclear feature.} & \parbox[t]{0.25\textwidth}{No clear large-scale bar; the nuclear feature remains ambiguous.} \\
\hline
\end{tabular}
\vspace{2pt}

\parbox{0.95\textwidth}{\raggedright Morphologies and mean family indices are from \citet{buta_classical_2015}; the NGC\,1317 notation is from \citet{sanchez_alarcon_cs4g_2025}. CGS velocities, distances, Galactic latitudes, and compiled optical types are from \citet{ho_carnegie-irvine_2011}. The S$^4$G parent-sample limits are from \citet{sheth_s4g_2010}.}
}
\end{table}

This comparison indicates that the control sample cannot be regarded as bar-free at all wavelengths. Several overlaps show evidence of a weak, intermediate, or nuclear bar. For NGC\,245, NGC\,5324, NGC\,6699, and NGC\,6935, we did not identify a homogeneous published NIR morphology classification. We retain all 15 galaxies because the calibration benchmark is defined by the uniform CGS optical classification. Excluding objects post hoc on the basis of heterogeneous longer-wavelength evidence would change that benchmark. Accordingly, the reported false-positive rates are interpreted relative to the CGS optical classification.

\FloatBarrier

\end{document}